\documentclass[sigconf, nonacm]{acmart}

\makeatletter
\def\@ACM@checkaffil{% Only warnings
    \if@ACM@instpresent\else
    \ClassWarningNoLine{\@classname}{No institution present for an affiliation}%
    \fi
    \if@ACM@citypresent\else
    \ClassWarningNoLine{\@classname}{No city present for an affiliation}%
    \fi
    \if@ACM@countrypresent\else
        \ClassWarningNoLine{\@classname}{No country present for an affiliation}%
    \fi
}
\makeatother

\newcommand\vldbdoi{XX.XX/XXX.XX}
\newcommand\vldbpages{XXX-XXX}
\newcommand\vldbvolume{18}
\newcommand\vldbissue{1}
\newcommand\vldbyear{2025}
\newcommand\vldbauthors{\authors}
\newcommand\vldbtitle{\shorttitle} 
\newcommand\vldbavailabilityurl{https://github.com/Fannxy/GORAM-ABY3}
\newcommand\vldbpagestyle{plain} 

\usepackage[linesnumbered, ruled]{algorithm2e}
\usepackage{enumitem}
\usepackage{amsmath,amsbsy,amsfonts}
\usepackage{sansmath}
\usepackage{textcomp}
\usepackage{etoolbox}
\usepackage{graphicx}
\usepackage{subfigure}
\usepackage{flushend}
\usepackage{hyperref}
\usepackage{url}
\usepackage{booktabs}
\usepackage{multirow}
\usepackage{threeparttable}
\usepackage{diagbox}
\usepackage{float}
\usepackage{setspace}
\usepackage{ulem}
\usepackage{caption}
\usepackage{stmaryrd}
\usepackage{xspace}
\usepackage{pifont} 
\usepackage{xcolor}
\usepackage{soul}
\usepackage{makecell}
\usepackage{tikz}

\usepackage[subject={Top1}]{pdfcomment}

\definecolor{lightgray}{RGB}{210, 210, 210}
\usepackage{newfloat}
\DeclareFloatingEnvironment[fileext=los,name=Protocol, placement={t},within=none]{Protocol}

\newcommand{\psec}[1]{{#1}}

\newcommand{\secret}[1]{$\llbracket #1 \rrbracket$}
\newcommand{\mathsecret}[1]{\llbracket #1 \rrbracket}
\sethlcolor{lightgray}  
\newcommand{\highl}[1]{\hl{#1}}

\newcommand{\code}[1]{{\small\texttt{#1}}}

\newcommand{\prot}[1]{{\small\sf{#1}}}
\newcommand{\para}[1]{\medskip\noindent\textbf{#1}}

\newcommand{\tinypara}[1]{\noindent\textbf{#1}}

\newcommand{\etal}{et al.\xspace} % yjp: 这里从空格改成了 \xspace，要不然 \etal 后面是空格或者 ~ 的时候会空特别大
\newcommand{\eg}{e.g.,\xspace}
\newcommand{\ie}{i.e.,\xspace}
\newcommand{\etc}{etc.\xspace}

\newcommand{\xor}{\mathbin{\oplus}}  

\newcommand{\gb}{G.{\small \sf{Block}}}
\newcommand{\eoram}{{\small \sf{EORAM}}\xspace}
\newcommand{\voram}{{\small \sf{VORAM}}\xspace}
\newcommand{\permil}{\text{\textperthousand}}

\renewcommand{\emph}[1]{{\textit{#1}}}

\newtheorem{theorem}{Theorem}

\newcommand{\titlesysname}{{\sf{GORAM}}\xspace}
\newcommand{\sysname}{{\small \sf \titlesysname}\xspace}

\newcommand{\mat}{{\emph{Mat}}\xspace}
\newcommand{\elist}{{\emph{List}}\xspace}

\SetKwComment{comm}{\rmfamily{\textit{\textcolor{gray}{// }}}}{}
\newcommand{\cfont}[1]{\rmfamily{\textit{\textcolor{gray}{#1}}}}

\begin{document}

\sloppy

\title{GORAM: Graph-oriented ORAM for Efficient Ego-centric Queries on Federated Graphs} % TODO: replace with your title

%  \author{
%  {\rm Anonymous Author(s)}\\
% }

\author{Xiaoyu Fan}
\affiliation{Tsinghua University}
\email{fxy23@mails.tsinghua.edu.cn}

\author{Kun Chen}
\affiliation{Ant Group}
\email{ck413941@antgroup.com}

\author{Jiping Yu}
\affiliation{Tsinghua University}
\email{yjp19@mails.tsinghua.edu.cn}

\author{Xiaowei Zhu}
\affiliation{Ant Group}
\email{robert.zxw@antgroup.com}

\author{Yunyi Chen}
\affiliation{Tsinghua University}
\email{cyy23@mails.tsinghua.edu.cn}

\author{Huanchen Zhang}
\affiliation{Tsinghua University}
\email{huanchen@tsinghua.edu.cn}

\author{Wei Xu}
\affiliation{Tsinghua University}
\email{weixu@tsinghua.edu.cn}

% \maketitle

\begin{abstract}
\emph{Ego-centric} queries, focusing on a target vertex and its direct neighbors, are essential for various applications. Enabling such queries on graphs owned by mutually distrustful data providers without breaching privacy holds promise for more comprehensive results.

In this paper, we propose \sysname, a graph-oriented data structure that enables efficient ego-centric queries on federated graphs with strong privacy guarantees. \sysname leverages \emph{secure multi-party computation (MPC)} and ensures that no information about the graphs or the querying keys is exposed during the process. 
For practical performance, \sysname partitions the federated graph and constructs an \emph{Oblivious RAM (ORAM)}-inspired index atop these partitions. This design enables each ego-centric query to process only a single partition, which can be accessed fast and securely.

Utilizing \sysname, we develop a prototype querying engine on a real-world MPC framework. We then conduct a comprehensive evaluation using five commonly used queries similar to the LinkBench workload description~\cite{linkbench2013} on both synthetic and real-world graphs. Our evaluation shows that all five queries can be completed in just 58.1 milliseconds to 35.7 seconds, even on graphs with up to 41.6 million vertices and 1.4 billion edges.
To the best of our knowledge, this represents the first instance of processing billion-scale graphs with practical performance on MPC.
\end{abstract}

\maketitle

%%% do not modify the following VLDB block %%
%%% VLDB block start %%%
\pagestyle{\vldbpagestyle}
\begingroup\small\noindent\raggedright\textbf{PVLDB Reference Format:}\\
\vldbauthors. \vldbtitle. PVLDB, \vldbvolume(\vldbissue): \vldbpages, \vldbyear.\\
\href{https://doi.org/\vldbdoi}{doi:\vldbdoi}
\endgroup
\begingroup
\renewcommand\thefootnote{}\footnote{\noindent
This work is licensed under the Creative Commons BY-NC-ND 4.0 International License. Visit \url{https://creativecommons.org/licenses/by-nc-nd/4.0/} to view a copy of this license. For any use beyond those covered by this license, obtain permission by emailing \href{mailto:info@vldb.org}{info@vldb.org}. Copyright is held by the owner/author(s). Publication rights licensed to the VLDB Endowment. \\
\raggedright Proceedings of the VLDB Endowment, Vol. \vldbvolume, No. \vldbissue\ %
ISSN 2150-8097. \\
\href{https://doi.org/\vldbdoi}{doi:\vldbdoi} \\
}\addtocounter{footnote}{-1}\endgroup
%%% VLDB block end %%%

%%% do not modify the following VLDB block %%
%%% VLDB block start %%%
\ifdefempty{\vldbavailabilityurl}{}{
\vspace{.3cm}
\begingroup\small\noindent\raggedright\textbf{PVLDB Artifact Availability:}\\
The source code, data, and/or other artifacts have been made available at \url{\vldbavailabilityurl}.
\endgroup
}
%%% VLDB block end %%%

\section{Introduction}\label{sec:intro}

Privacy-preserving federated query engines allow data providers to collaboratively compute the query result while keeping the input and all the intermediate results private.  It becomes increasingly important with demands from many industries to share their data. 
% However, existing engines either limit the schema to tables or face severe scalability limitations. 

% background on the importance of ego-centric queries over the federation of graphs.
The purpose of this paper is to support federated queries on graphs, specifically the crucial \emph{ego-centric} queries that target a specified vertex and all its direct neighbors. 
This basic type of queries has many applications, especially if multiple parties can cooperate. 
For example, by analyzing the relations among suspicious accounts, commercial banks can effectively detect money laundering from the transaction graphs~\cite{neo4financial, qiu2018real}. Quick identification of people exposed to an infected person is essential in a pandemic, and obviously the results become more thorough if we join data from multiple providers~\cite{liu2024making}. As LinkBench~\cite{linkbench2013} reports, neighbor filtering on the given account, a typical ego-centric query, constitutes 55.6\% of all queries at Meta. In fact, all queries in LinkBench are ego-centric.

The above motivating examples demonstrate that both the graph and the query keys contain sensitive information. Therefore, it is crucial to keep both of them private. While ego-centric queries are easy to do in plaintext, their efficiency highly depends on the graph-data-aware organizations, such as caches for popular vertices~\cite{yang2021large, bronson2013tao}. However, this is in conflict with the privacy requirements in our federated setting.
Unlike tabular data, where the schema is public and we only need to protect the data and query keys, the \emph{graph structure} information also needs to be protected.  E.g., whether a particular vertex or edge exists, the degree of a certain vertex, and even the distribution of the vertex degrees are all considered private information. 
For example, disclosing an edge in the transaction graph could expose sensitive relationships between accounts, thus necessitating protection.
% Research has shown severe privacy breaches using graph structure~\cite{}\todo{}.

One typical method to implement private queries on federated data is to use \emph{secure multi-party computation (MPC)}~\cite{4568207} throughout the query process~\cite{liagouris2023secrecy,bater2017smcql,volgushev2019conclave,bogdanov2008sharemind,asharov2023secure}.
MPC is a cryptographic technique that allows multiple parties to jointly compute a function on their private inputs, guaranteeing that no information is leaked.

The straightforward idea for graph is to encode the entire adjacency matrix with MPC~\cite{blanton2013data} and thus hide the entire graph, including the structure. 
Unfortunately, the representation requires at least $O(|V|^2)$ space, where $|V|$ is the number of vertices.  As real-world large graphs are sparse~\cite{chung2010graph}, this representation wastes a vast amount of memory to store non-existent edges (to protect the existence of edges), making it impractical for large graphs.
% prohibitively expensive for larger graphs. 

Nayak \etal~\cite{nayak2015graphsc} propose an innovative way to protect the graph structure by encoding and encrypting both the vertices and edges into an identical form and storing them in a \emph{secure list}, thereby reducing the space overhead to $O(|V|+|E|)$, where $|E|$ is the number of edges. 
Given that $|V|+|E| \ll |V|^2$ for real-world graphs, the efficient space utilization makes this representation popular in secure graph processing~\cite{nayak2015graphsc,araki2021secure,mazloom2018secure,mazloom2020secure,koti2024mathsf}.
However, to perform an ego-centric query while keeping the query key private, we need to scan the entire list, leading to $O(|V|+|E|)$ time complexity. 

We observe that despite the space-inefficiency, using a well-studied cryptographic technique called \emph{Oblivious RAM (ORAM)}~\cite{goldreich1996software,zahur2016revisiting}, which allows accessing the $i$th element of an array in sublinear time without revealing $i$, we can secretly access any element of an adjacency matrix sublinearly, providing sublinear query time.
Our key idea is that if we can split the edge list into multiple partitions and build a matrix on top of the partitions, we can reduce the space size and only scan one partition for each query.

We propose \sysname, a graph-oriented data structure to support efficient ego-centric queries on large-scale federated graphs with strong privacy guarantees.
In a nutshell, \sysname splits the vertices into multiple chunks and then segments the graph into a “matrix” of edge lists. We carefully plan the segmentation so that each edge list contains all edges starting from vertices in the row’s chunk and destinations in the column’s.
\psec{In this way, the graph is organized as multiple \emph{partitions}, satisfying that all the information needed for each ego-centric query is contained in \emph{exactly} one partition.}
Logically, all partitions together contain the entire edge list of the graph, and the matrix serves as a secure \emph{index} mapping a vertex or edge ID to a particular partition.  Using \sysname, we can locate a particular partition sublinearly, and then we only need to scan the partition, greatly reducing running time (Section~\ref{subsec:cons_and_access}).
While current ORAM only supports addressing into a single element, we extend the idea so that we can access a full partition efficiently and privately, maximizing benefits from vectorization and parallelism (Section~\ref{subsec:parall}). 
While \sysname is designed agnostic to underlying MPC protocols, we find that some widely used protocols, \eg ABY3~\cite{mohassel2018aby3}, allow us to design a constant-round shuffling protocol, vastly accelerating ORAM initiation (Section~\ref{subsec:shuffmem}).
% what we have achieved and contributions.
Based on \sysname, we can easily implement crucial queries. 
We implement five types of ego-centric queries, including \emph{edge existence}, \emph{1-hop neighbors}, \emph{neighbors filtering}, \etc, covering all the LinkBench queries~\cite{linkbench2013}(Section~\ref{sec:local_query}).

We evaluate the above queries on three real-world and thirty synthetic graphs with varied distributions and sizes.
Results in Section~\ref{sec:eval} show remarkable efficiency and scalability of \sysname. On the largest graph, Twitter~\cite{twitter}, with more than 41.6 million vertices and 1.4 billion edges, all the queries can be finished within 35.7 seconds, and the fastest query only takes 58.1 milliseconds, showing 2 to 3 orders of magnitude speedup over the existing secure graph data structures.
Also, the initialization of \sysname only requires less than 3.0 minutes.
We provide a detailed performance analysis using the synthetic graphs and show that \sysname outperforms the existing data structures across varied distributions and sizes.
To the best of our knowledge, \sysname is the first to support graphs with more than one \emph{billion} edges in secure computations, which is 2-orders-of-magnitude larger than the prior arts~\cite{araki2021secure,mazloom2020secure,koti2024mathsf}.

In summary, our contributions include:

(1) We propose \sysname, a graph-oriented data structure to support efficient sublinear ego-centric queries on federated graphs, guaranteeing strong privacy.

(2) We design comprehensive optimizations for practical performance on large-scale graphs, including local processing, lifecycle parallelisms, and a constant-round shuffling protocol.

(3) We develop a prototype secure querying engine based on \sysname and evaluate it comprehensively using five commonly used queries on 33 synthetic and real-world graphs, demonstrating remarkable efficiency and scalability.

%3) A prototype querying engine based on \sysname is developed on a real-world MPC framework and evaluated comprehensively using five commonly used queries on more than thirty synthetic and real-world graphs, demonstrating remarkable efficiency and scalability.
\section{Cryptography Background}\label{sec:background}

\subsection{Secure Multi-party Computation (MPC)}\label{subsec:mpc}

Secure Multi-party Computation (MPC) allows multiple distrusting parties to jointly compute a function while keeping each party's input private. Note that the role of a ``party'' is to participate in the computation rather than protecting their private data.

% \tinypara{Secret sharing} is a popular technique in MPC~\cite{4568207}. A \emph{$(t,n)$-secret sharing} schema splits sensitive data $x$ to $n$ parties, satisfying that any $t$ parties can reconstruct $x$ while fewer than $t$ parties learn nothing about $x$.  Note that the ``parities'' refer to independent participants in the computation, which is different from ``data providers'' who use the computation system to protect their private data. 

% \sysname adopts the $(2,3)$-secret sharing for efficiency, similar to prior works~\cite{araki2021secure,liagouris2023secrecy,mohassel2018aby3,falk2023gigadoram}. 
% Primarily, \sysname adopts the \emph{boolean} secret share. In this scheme, a number $x$ is split into $(x_1, x_2, x_3)$, such that each $x_i, i \in \{1, 2, 3\}$ is uniformly random and $x \equiv x_1 \xor x_2 \xor x_3$, where $\xor$ is bitwise \code{XOR}. 
% Each party owns two shares $(x_i, x_{i+1})$ with indices wrapped around 3. Therefore, any party learns nothing about $x$ while any two parties can perfectly reconstruct $x$.
% We denote the boolean secret shares of $x$ as \secret{x}. 
% For efficient arithmetic operations, we transform \secret{x} into \emph{arithmetic} shares, \ie $x \equiv x_1 + x_2 + x_3 \pmod {2^k}$, denoted as $\mathsecret{x}^{A}$. 

\tinypara{Secret sharing} is a popular technique in MPC~\cite{4568207}. A \emph{$(t,n)$-secret sharing} schema splits sensitive data $x$ to $n$ parties, satisfying that any $t$ parties can reconstruct $x$ while fewer than $t$ parties learn nothing about $x$.
Similar to~\cite{araki2021secure,liagouris2023secrecy,mohassel2018aby3,falk2023gigadoram}, \sysname adopts the efficient $(2,3)$-secret sharing with \emph{boolean} secret share. This scheme splits a number $x$ into $(x_1, x_2, x_3)$ with each $x_i$ ($i \in \{1, 2, 3\}$) being uniformly random such that $x \equiv x_1 \xor x_2 \xor x_3$ ($\xor$ denotes bitwise \code{XOR}).
Each party owns two shares: $(x_1, x_2)$, $(x_2, x_3)$, or $(x_3, x_1)$.
Any particular party learns nothing about $x$, but any two parties can reconstruct $x$.
Denote the boolean secret shares $(x_1, x_2, x_3)$ as \secret{x}.

% \tinypara{Secure operations.}  With the secret shares \secret{x} and \secret{y}, the parties can collaboratively compute the shares of the result \secret{z} for a variety of operations (\code{op}) using MPC protocols (\prot{OP}), \eg \prot{XOR}, \prot{AND}, \prot{OR}, $+$, $\times$, comparisons($>, \geq, =$) and transformations between \secret{x} and $\mathsecret{x}^{A}$~\cite{demmler2015aby,mohassel2018aby3}. 
% The protocols guarantee both correctness, i.e., for $z = \text{\code{op}}(x, y)$, $\mathsecret{z} = \prot{OP}(\mathsecret{x},\mathsecret{y})$, and privacy, i.e., any parties learn nothing about the secret inputs. 
% % We can also compose these operations to design advanced algorithms~\cite{rathee2021sirnn, gupta2022llama, keller2022secure,liagouris2023secrecy,volgushev2019conclave,bater2017smcql}.
% % Composing these basic operations, people have designed high-level secure algorithms and data structures~\cite{rathee2021sirnn, gupta2022llama, keller2022secure}, including relational queries~\cite{liagouris2023secrecy,volgushev2019conclave,bater2017smcql}.
% Of course, privacy comes with a cost. Except for \prot{XOR}s on boolean shares and additions on arithmetic shares, all other operations require at least one round of communication among the computation parties. To amortize the communication overhead, it is popular to batch up multiple operations into a single vector, similar to SIMD~\cite{liagouris2023secrecy,mohassel2018aby3}.

\tinypara{Secure operations.} For a variety of operations (\code{op}) such as \prot{XOR}, \prot{AND}, \prot{OR}, $+$, $\times$, and comparisons ($>, \geq, =$), we can use the MPC protocols (\prot{OP}) to compute $z = \text{\code{op}}(x, y)$ collaboratively and securely: $\mathsecret{z} = \prot{OP}(\mathsecret{x},\mathsecret{y})$, where efficient arithmetic operations rely on transforming \secret{x} into $\mathsecret{x}^{A}$ such that $x \equiv x^{A}_1 + x^{A}_2 + x^{A}_3 \pmod {2^k}$~\cite{demmler2015aby,mohassel2018aby3}. Except for \prot{XOR} and $+$, other operations require at least one round of communication among the computation parties. It is common to batch the operations (similar to SIMD) to amortize the communication cost~\cite{liagouris2023secrecy,mohassel2018aby3}.
% \hz{Huanche stopped here.}

% In the case of only three parties, honest-majority is equivalent to \emph{non-colluding}, \ie only one party can be corrupted. 

\tinypara{Security guarantees.} A well-designed MPC protocol can guarantee privacy and correctness against a specific \emph{adversary model}.
Common classifications include \emph{semi-honest} vs. \emph{malicious}, determined by whether the corrupted parties can deviate from the protocol and \emph{honest-} vs. \emph{dishonest-} majority, contingent on the corruption proportion. 
\sysname adopts all its MPC protocols from ABY3~\cite{mohassel2018aby3} and Araki \etal~\cite{araki2021secure}, thus inheriting their semi-honest and honest-majority adversary model, same as~\cite{araki2021secure,liagouris2023secrecy,mohassel2018aby3,falk2023gigadoram}.
Note that \sysname is agnostic to the underlying protocols and can be adapted to others.
% (may lead to different security and performance). 

\subsection{Oblivious RAM (ORAM)}\label{subsec:oram}

\tinypara{Oblivious RAM (ORAM)}~\cite{goldreich1996software} aims to implement oblivious \emph{indexing}, \ie accessing the $i$th element in an array while keeping $i$ secret. ORAM offers two desirable properties: 1) it hides the access patterns, \ie for \emph{any} two indices $i, j$, servers performing the access cannot distinguish whether the index is $i$ or $j$, protecting the privacy of the query keys;
2) it enables sub-linear-complexity accesses, thereby providing scalability. 
The original ORAM, \eg Path ORAM~\cite{stefanov2018path}, is designed for the \emph{client-server} scenario, where a single client stores and retrieves her own private data on a single untrusted server~\cite{goldreich1996software}. In other words, ORAMs do not support the case where the underlying data comes from multiple data providers nor allow a third-party client (not the data provider) to access data.

\tinypara{Distributed ORAM (DORAM)} adds support to multiple data providers by using a group of computation servers that together hold the secret-shared data~\cite{lu2013distributed}. The servers can jointly access the target secret element given the secret-shared index, \ie compute \secret{arr[i]} $ = \mathsecret{arr}[\mathsecret{i}]$ without leaking the index.
% Efficient DORAM implementations include~\cite{falk2023gigadoram, vadapalli2023duoram, bunn2020efficient, zahur2016revisiting, doerner2017scaling}.
We build the index of \sysname by extending the classic Square-root ORAM~\cite{zahur2016revisiting}. 
% The key idea of Square-root ORAM is to shuffle the original data ($arr$) according to a random permutation $\pi$, then build an index map on $\pi$ that translates the secret logical index $\mathsecret{i}$ to the plaintext physical index $p$ in the shuffled data, satisfying that $arr[i] \equiv \widetilde{arr}[p]$. Also, the translation procedure is \emph{oblivious}, \ie the access pattern is independent with the input index, therefore keeping $i$ secret.
Note that \sysname is agnostic to the specific DORAM implementations, and we choose Square-root ORAM because it has two desirable properties, in addition to simplicity: (1) it offers sublinear access time, and (2) it circumvents the reliance on any specific ciphers that may potentially degrade security. Section~\ref{sec:rela_work} provides discussion on other DORAM structures~\cite{falk2023gigadoram, vadapalli2023duoram, bunn2020efficient, zahur2016revisiting, doerner2017scaling}. All the ORAM in the following denotes Square-root ORAM.

\section{Overview}\label{sec:overview}

\begin{figure}[t]
    \centering
    \includegraphics[width=0.38\textwidth]{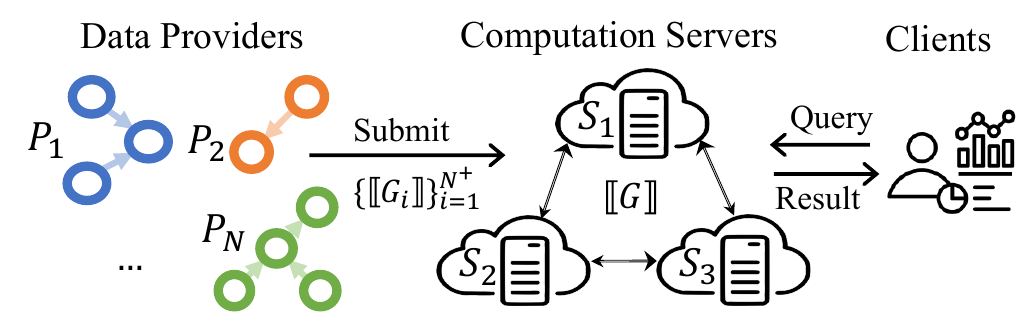}
    \vspace{-.1in}
    \caption{Logical Roles in Private Querying Process}
    \label{fig:roles}
    \vspace{-.15in}
\end{figure}

\sysname is a query engine designed to support private, ego-centric queries on federated graphs.  There are three types of roles:  
(1) arbitrary number of \emph{data providers}, each of which owns part of the graph that we call the \emph{private graph};
(2) three \emph{computation servers} that run the MPC protocol to process queries; 
and (3) an arbitrary number of \emph{clients} who submit queries to the computation servers and receive the results. 
Similar to \cite{liagouris2023secrecy}, the roles are decoupled. Each party can hold any combination of \emph{different} roles, \eg it can be both a data provider and a server.
To be useful in the real world, we want \sysname to satisfy the following requirements: 

\emph{(1) Functionality:} we want to support \emph{arbitrary} ego-centric queries, allowing us to perform any filter or aggregation on a target vertex or edge along with all its direct neighbors.

\emph{(2) Privacy:} we want to keep two information private: (a) the query keys, \ie the target vertices or edges; and (b) the graph, including the graph structure and the attributes of vertices and edges. 
Also, although the query result is revealed to the client, the client cannot infer which data provider contributed to the result.

\emph{(3) Scalability:} we want to support graphs with \emph{billions} of edges, meeting the need for real-world graphs.  Also, we anticipate responses within a few seconds on graphs with this scale.

\subsection{Formalization}\label{subsec:formalization}

% We first provide the formal definition and notations. 

\tinypara{Private ego-centric queries on federated graphs. } We assume a \emph{global} directed graph $G=(V, E)$ is distributed among $N$ data providers $P_i, i \in [N]$, $[N]=\{1, 2, ..., N\}$.  
$V$ and $E$ denote the vertex and edge sets, respectively. 
Each edge $e \in E$ has a source and a destination vertex, $v_s$ and $v_d$. 
We assume the edges $(v_s, v_d)$ are different because they may contain different attributes, \eg timestamps~\cite{linkbench2013}. We transform undirected graphs into directed ones by representing each edge with two directed edges.

Each \emph{data provider} $P_i$ owns a private graph $G_i = (V_i, E_i)$, satisfying that $V_i \subseteq V$, $E_i \subseteq E$. $E = \cup_{i\in[N]} E_i$ is the set of all edges, and it is always private. 
$V$ is the set of all possible vertices, which is public. 
Because the complexity of \sysname is independent of $|V|$, we can safely set it to all the possible vertex names (\eg a 64-bit ID), without leaking any information about the actual set of vertices.  
For \mat that does not support a large $V$, we set $V$ as the union of all $V_i$'s and make it public.
There are three semi-honest \emph{computation servers} $S_i, \, i \in \{1, 2, 3\}$ holding the secret global graph \secret{G} $= (V, \mathsecret{E^{+}})$ and carrying all the query processing, where $E^{+} \supseteq E$ is a super-set of $E$ because it may contain dummy edges for privacy.   
Each \emph{client} can submit an ego-centric query with a secret key of either a vertex $\mathsecret{v}$ or edge $(\mathsecret{v_s}, \mathsecret{v_d})$ ($v, v_s, v_d \in V$) to the computation servers and receive the results. The ego-centric query can be an arbitrary filter or aggregation on the sub-graph $G_{\text{sub}} = (V_{\text{sub}}, E_{\text{sub}})$ containing all the direct neighbors of the target vertex $v$ and the corresponding edges $(v, v^{*})$ if $(v, v^{*}) \in E$, or all edges $(v_s, v_d) \in E$. Ego-centric queries refer to both vertex- and edge-centric queries.
% Both vertex- and edge-centric queries as ego-centric queries.

\tinypara{Threat model.} We assume all roles \emph{semi-honest}, and there exists an adversary who can compromise at most one computing server and see all of its internal states, similar to~\cite{liagouris2023secrecy, asharov2023secure,araki2021secure,falk2023gigadoram}.
Also, we assume the authentication of the data providers is conducted through the \emph{ring signature}~\cite{noether2016ring}, a cryptographic technique enabling anonymous authentication.

% I.e., all parties can verify the authentication of the data providers without revealing their identities.

% data providers are semi-honest and their identities are authorized through the \emph{ring signature}~\cite{noether2016ring}, which is used for anonymous authentication, i.e., all the parties can verify whether the data provider is authorized or not while can not tell who it is.}

\begin{figure*}[t]  
    \centering  
    \begin{minipage}[t]{0.63\textwidth}  
        \centering  
        \includegraphics[width=\textwidth]{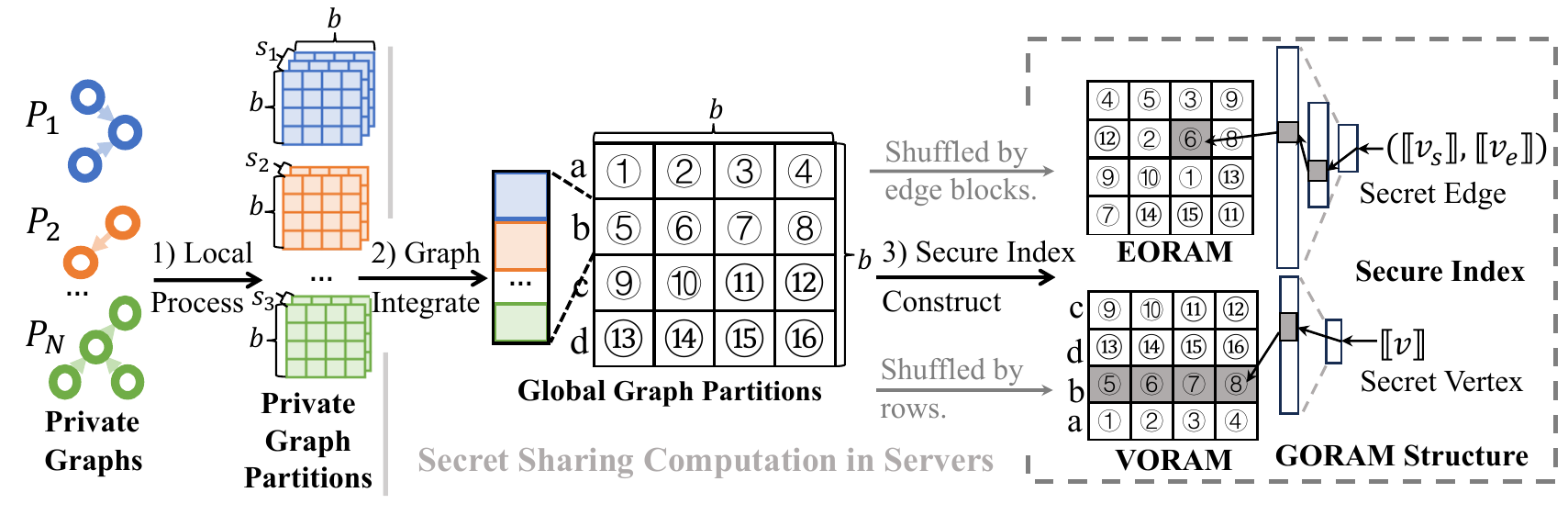}  
        \vspace{-.2in}
        \caption{\sysname Initialization and Structure Overview}  
        \label{fig:goram_overview}  
    \end{minipage}  
    % \hfill  
    \begin{minipage}[t]{0.33\textwidth}  
        \centering  
        \includegraphics[width=\textwidth]{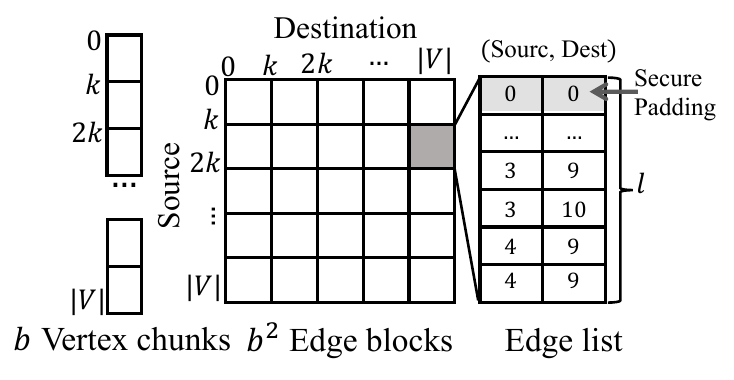} 
        \vspace{-.2in} 
        \caption{Graph Partition ($k=2, |V|=10$)}  
        \label{fig:2d_partition}  
    \end{minipage}  
    \vspace{-.13in}
\end{figure*}

\subsection{Strawman Solutions}\label{subsec:motivation}

Two classic data structures are used to present the secure graph, \ie \emph{adjacency matrix} (\mat)~\cite{blanton2013data} and \emph{edge list} (\elist)~\cite{araki2021secure,nayak2015graphsc,mazloom2018secure,mazloom2020secure}.

\tinypara{Based on \mat. } With the public vertex set $V$, data providers can locally construct the $|V|^2$ adjacency matrix, encrypt it into secret-shared matrix and transfer the shares to the computation servers. 
Each server adds up the $N$ secret matrices to form the secret matrix of the global graph $G$. 
Because \mat is a $|V| \times |V|$ matrix, we can naively adopt the ORAM for efficient access. Specifically, we build two ORAMs, adj-\voram, operating over $|V|$ matrix rows (source vertices) and adj-\eoram, operating over an array of $|V|^2$ matrix elements (edges). 
For a vertex query on $v_i$, we access the $\mathsecret{i}$th row from the adj-\voram. This row contains $|V|$ elements, each representing the edge number between $v_i$ and $v_j, j \in [|V|]$.  For an edge query on $(v_i, v_j)$, we access the element $\mathsecret{i*|V|+j}$ from the adj-\eoram, which contains the number of edge $(v_i, v_j)$.

\mat is simple but never practical because of three issues. 
The most fatal one is the $O(|V|^2)$ space cost, making it impractical for real-world sparse graphs~\cite{linkbench2013,slashdot,dblp, chung2010graph}.  
Secondly, to limit $V$, we need to expose the actual vertex set, instead of just the possible namespace, public to all parties, leaking the information of the global set (\eg the global customer list).  
Lastly, we cannot support multi-graph, \ie edges with different attributes between a single pair of vertices, which is desirable in applications like transactions.

%Also, we can not support the cases that the edges with the same start and end vertices contain different attributes. To support these cases directly, data providers can place multiple edges $(v_i, v_j)$ with different attributes into the $(i, j)$th matrix cell. However, this would compromise privacy as the servers could infer which edge is more prevalent in the private graphs. To protect this information, we must pad each cell to align with the maximum one using dummy edges, which exacerbates the already considerable space cost.

\tinypara{Based on \elist.} Each data provider $P_i$ can simply create a secret-shared list of its edges $(\mathsecret{u}, \mathsecret{v})$, and optionally other attributes like timestamps.  The compute servers can then concatenate the $N$ secret lists to create the global edge list.
In this way, the only information leaked to the servers is the total number of edges from each provider. 
To protect these numbers, providers can append extra $\epsilon_i$ dummy edges, \ie $(\mathsecret{0}, \mathsecret{0})$, before sending to the servers. 
% The sorted order is beneficial for some queries, \eg \code{NeighborsGet}, which will be discussed in Section~\ref{sec:local_query}.

Although edge list is much more compact, the main issue with it is the access efficiency.  
Because all the edges are encrypted, and we have no knowledge about the existing vertices and edges, we can only scan the whole \elist for each ego-centric query, which introduces $O(|E|)$ complexity.

\tinypara{In summary, } the two strawman solutions are either too costly on storage or on access time.  A viable solution should allow a compact space while supporting sublinear access complexity. 

%methods contain theoretical limitations to satisfy the practical requirements. \elist requires scanning the whole edge list even if the query targets a single vertex/edge only, limiting the performance; \mat introduces impractical space cost and can not support arbitrary ego-centric queries.

% \begin{figure*}[t]  
%     \centering  
%     \begin{minipage}[t]{0.63\textwidth}  
%         \centering  
%         \includegraphics[width=\textwidth]{./graph/GORAM_initialization_and_structure.pdf}  
%         \vspace{-.18in}
%         \caption{\sysname Initialization and Structure Overview}  
%         \label{fig:goram_overview}  
%     \end{minipage}  
%     % \hfill  
%     \begin{minipage}[t]{0.33\textwidth}  
%         \centering  
%         \includegraphics[width=\textwidth]{./graph/2d_partition.pdf} 
%         \vspace{-.18in} 
%         \caption{Graph Partition ($k=2, |V|=10$)}  
%         \label{fig:2d_partition}  
%     \end{minipage}  
%     \vspace{-.1in}
% \end{figure*}

\subsection{\titlesysname Design}\label{subsec:GORAM_over}

%\tinypara{Motivation and key idea.} The above two data structures actually represent the graph in two extreme ways: \elist is space-efficient but slow for queries, while \mat is query-efficient but introduces impractical space cost. 
%This contrast motivates us to design a data structure that strikes a balance between the two extremes and leverages the advantages of both. 

The high-level idea of \sysname is to split the graph into a ``matrix'' of edge lists. 
Intuitively, the matrix structure enables building ORAMs on top of the graph, circumventing the need for a full scan for each query. Simultaneously, the use of internal edge lists averts the $O(|V|^2)$ space complexity associated with the \mat structure. This approach seeks to achieve both space- and query-efficiency.

\tinypara{\sysname Overview. }
As Figure~\ref{fig:goram_overview} shows, \sysname is a secret-shared data structure of the global graph \secret{G}, held by the computation servers. 
\sysname randomly shuffles the public vertex set $V$, splits it into $b$ vertex chunks and then splits the graph \secret{G} into a ``matrix'' of $b^2$ blocks, each block contains \emph{all} the edges starting from and ending in two specific vertex chunks, and each row of the blocks correspondingly contains \emph{all} the direct neighbors of a specific vertex chunk. This gives us two types of \emph{partition}s: the block forms the partition for edge-centric queries, and the row of the blocks constitutes the partition for vertex-centric queries.
\sysname then constructs \secret{G} into \voram and \eoram, which can securely access the partition given the secret vertex or edge. 
The indexing is achieved by modeling the partitions as ORAM, and we extend its functionality from accessing \emph{array-of-elements} to \emph{array-of-partitions}. 

\sysname can be securely and efficiently initialized through three steps: 
(1) each data provider locally processes her private graph into secret-shared partitions;
(2) the computation servers integrate all the partitions of private graphs into the global graph \secret{G};
and (3) the computation servers construct the secure indices for the partitions. The details of \sysname are illustrated in Section~\ref{sec:engine}.

Through \sysname, we can implement arbitrary and efficient ego-centric queries easily. 
The clients can submit a provided query with a secret query key. 
For each query, the computation servers receive the query key, find the partition containing the key using \sysname, execute the query on the partition, and return the result to the client.
We provide five query examples in Section~\ref{sec:local_query}.

% The complexities of \sysname and the other two strawman solutions are summarized in Table~\ref{tab:complexity}, and Section~\ref{subsec:eval-performance} provides the detailed performance evaluation comparing \sysname with the strawman solutions. 

% and the user-defined function (e.g. an aggregation) from the client, find the partition containing the key using \sysname, run the user-defined function on the partition, and return the result to the client. 

% \hz{The above description is a bit confusing at first because you implicitly assume that the graph is represented as an adjacency matrix. State that up front and define ``vertex groups' first.}

\tinypara{Privacy guarantees. } \sysname provides two guarantees:
(1) \emph{clients' query key privacy}: no other party can learn anything about the client's query key;
(2) \emph{data providers' graph privacy}: no other party can learn any information about the data provider's private graph, which includes the graph structure and the attributes of vertices and edges. Additionally, the client who receives the query result learns nothing except the result, including which provider contributes to it. Section~\ref{subsec:security} shows how \sysname ensures these guarantees.

\section{Graph-Oriented ORAM (\titlesysname)}\label{sec:engine}

To satisfy the requirements in Section~\ref{sec:overview}, \sysname splits the global graph $G$ into a ``matrix'' of edge lists, forming a 2d-partitioned data structure, as Figure~\ref{fig:2d_partition} shows.
This structure groups every successive $k$ vertices into a chunk according to randomly shuffled IDs from the range $[|V|] = \{1,2,\dots,|V|\}$, thereby creating $b = \lceil \frac{|V|}{k} \rceil$ chunks.
Then, it splits the global graph into $b^2$ blocks of edge lists for each pair of chunks.
Specifically, each block $(s, d)$ contains all the edges $\{(v_s, v_d)\}$, with $v_s$ and $v_d$ belonging to the $s$th and $d$th chunk, respectively. 
\psec{
% Furthermore, to maintain security during the querying process, each block is made equal in length by being filled with dummy edges. This ensures that the blocks remain indistinguishable throughout the secure computation.
To ensure security, each block is equalized in length with dummy edges, making the blocks indistinguishable.
}

It is worth noting that if an edge $(v_s, v_d)$ exists, it is contained in a single edge block $(\lceil \frac{v_s}{k} \rceil, \lceil \frac{v_d}{k} \rceil)$. For each vertex $v$, all the direct outing neighbors\footnote{If the bi-directional neighbors are interested, the in-going neighbors are included in the $\lceil \frac{v}{k} \rceil$th column of the 2d-partition.} are included in the $\lceil \frac{v}{k} \rceil$th row of $b$ blocks.
We heuristically choose a $k$ based on the density of the graph because many real-world graphs like social graphs or transaction graphs have well-known character $D$~\cite{chung2010graph}.  Given that $D=\frac{|E|}{|V|}$ is the density of the graph, we use a default $k = \frac{|V|}{D}$, making the $b$ independent of $|V|$.
The block and row of blocks are the graph \emph{partitions} for edge- and vertex-centric queries, respectively. 
Obviously, \emph{only} one partition needs to be processed for each query.
\sysname then builds the partitions as ORAMs to enable secure and fast access.

% In practice, we do not want $k$ to be too large, as it leads to scans over large blocks, reducing the effectiveness of the ORAM index.  However, $k$ cannot be too small either, as we still need to hold enough edges in each block to facilitate vectorization.

\subsection{Preliminaries of Square-root ORAM}\label{subsec:goram-pre}

% Before delving into the details of \sysname, we provide a brief introduction to Square-root ORAM~\cite{zahur2016revisiting}, which is extended and optimized by \sysname to support graph queries and more efficient initialization.

For readers unfamiliar with this area, we briefly introduce the Square-root ORAM~\cite{zahur2016revisiting}, which \sysname extends for graph queries.
% Before detailing \sysname, we briefly introduce the Square-root ORAM~\cite{zahur2016revisiting}, which \sysname extends for efficient graph queries.

The key idea of Square-root ORAM is to shuffle an array ($arr$) according to a random permutation $\pi$, then build an index map on $\pi$ that translates the secret logical index $\mathsecret{i}$ to the plaintext physical index $p$ in the shuffled data ($\widetilde{arr}$), where $arr[i] \equiv \widetilde{arr}[p]$. 
Specifically, after shuffling $\widetilde{arr} = \pi(arr)$, ORAM explicitly stores the permutation representation $\vec{\pi} = \pi^{-1}(L), L = \{0, 1, \dots n-1\}$ in secret shares, where $n$ is size of $arr$ and $\pi^{-1}$ is the inverse permutation of $\pi$, \ie $\pi^{-1}(\pi(arr)) \equiv arr$.
Each $\vec{\pi}[i]$ records the location of $arr[i]$ in $\widetilde{arr}$. The index map of $\pi$ is constructed as a recursive ORAM, where $P$ successive elements of $\vec{\pi}$ are packed as a single element and then build an ORAM for the $\frac{|\vec{\pi}|}{P}$ elements recursively until the size is no greater than $T$.
To access the $j$th element in $arr$ (\ie $arr[j]$), it is equivalent to access the $(\vec{\pi}[j])$th element in $\widetilde{arr}[\vec{\pi}[j]]$. Then we recur to the ORAMs built for $\vec{\pi}$ to find $\vec{\pi}[j]$.

Intuitively, each different logical index \secret{i} reveals a different random index, thereby keeping the access pattern private. To address the issue when repeating the same logical indices, Square-root ORAM employs a \emph{stash} that stores the accessed element each time. For each logical index, we first try to find it in the stash. We only use this index to access the ORAM if it is \emph{not} in the stash; otherwise, we pick an unused index to access the ORAM.  Therefore, one always sees an access to stash and one to the ORAM and thus cannot distinguish whether the logical index is repeated. 
When the stash reaches capacity, we need to rebuild the ORAM. By default, the stash size is set to $T = \sqrt{n}$, where $n$ is the number of elements in the ORAM.
For read-only accesses, ORAMs can be continuously built in the background, allowing immediate replacement with a fresh ORAM when the stash is full. Disregarding the cost of rebuilding, the average access cost over $T$ elements is $O(PT\log(\frac{n}{T}))$.

\subsection{\titlesysname Initialization and Access}\label{subsec:cons_and_access}

% We now introduce the \sysname initialization and partition access.
The \sysname initialization involves three steps:

\tinypara{Step 1 - Each data provider locally partition the private graph. }
Because the global $V$ is public, each data provider $P_i$ can locally split the private graph $G_i$ into the same $b^2$ blocks according to the public $k$.
Each $P_i$ initializes $b^2$ blocks using the global $V$ regardless of the vertices actually owned in the private $V_i$.
$P_i$ then traverses the edge list $E_i$, pushing each edge $(v_s, v_d)$ to block $(s, d)$, where $\lceil \frac{v_s}{k} \rceil = s$ and $\lceil \frac{v_d}{k} \rceil = d$, and sorting each block using the key $v_s||v_d$ (\ie concatenation of the source and destination vertices).\footnote{A sorted order is beneficial for some queries, as discussed in Section~\ref{sec:local_query}.}

To protect the graph structure, it's crucial that each block is aligned to a uniform size $\bar{l}$. This public parameter is heuristically set to a default value of $8$ to avoid excessive padding. 
% Without alignment, servers can infer at least two private information: 1) which chunks of vertices have more edges, and 2) the difference among the data providers, \eg one has vertices with more edges than others. 
The alignment includes two steps:
(1) Each $P_i$ pads blocks with dummy edges, \ie $(0,0)$, to align all blocks to the maximum block size $l_i$, thus protecting the edge distributions of each chunk; 
(2) If $l_i$ is less than $\bar{l}$, $P_i$ pads each block with $\bar{l} - l_i$ dummy edges. Otherwise, $P_i$ divides the partitioned graph into $s_i = \lceil \frac{l_i}{\bar{l}} \rceil$ sub-partitions. Each sub-partition contains $b^2$ blocks, and each block holds $\bar{l}$ edges, protecting the variance among the data providers.

\tinypara{Step 2 - MPC servers globally integrate the private partitions. } 
After dividing the private graph $G_i$ into $s_i$ $(b \times b \times \bar{l})$ sub-partitions, $P_i$ encrypts all elements, including all IDs and attributes, into boolean secret shares and sends the shares to the computation servers. Each encrypted and transferred partition is a $(b \times b \times \bar{l})$ secret matrix and is indistinguishable from the others, we refer to it as $\mathsecret{G_{u}.{\sf{Block}}}$.

The data providers send each $\mathsecret{G_{u}.{\sf{Block}}}$ through a separate anonymous channel, authorized via the \emph{ring signature}~\cite{noether2016ring}. This ensures that the computation servers only receive $N^{+} = \sum_{i=1}^{i=N}s_i$ secret $\mathsecret{G_{u}.{\sf{Block}}}$s, but cannot identify which one comes from which data provider.
The servers then run $b^2$ secure \emph{odd\_even\_merge\_sort} networks concurrently to merge blocks from each $\mathsecret{G_{u}.{\sf{Block}}}$, thereby constructing the global secret partitioned graph, $\mathsecret{\gb}$. The size of $\mathsecret{{\gb}}$ is ($b \times b \times l$), where $l = \sum_{i=1}^{N^{+}}\bar{l}$.

\tinypara{Step 3 - MPC servers construct secure index.} 
To enable secure and efficient access to the target partition for each query, \sysname builds the partitions as ORAMs.
Specifically, we model the partitioned graph $\mathsecret{\gb}$ with two ORAMs for vertex- and edge-centric queries: 
(1) \voram models $\mathsecret{\gb}$ as an array (\code{Arr}) of $b$ partitions, each is one row of $b$ blocks; and 
(2) \eoram models $\mathsecret{\gb}$ as an array of $b^2$ partitions, each is an edge block. The indices of \eoram are the flattened indices of the blocks, \ie the index of block $(i, j)$ is $ib + j$. 

The two sub-ORAMs are initialized in the following way, similar to ORAM:
(1) shuffling the graph in the unit of partitions according to a random permutation $\pi$ to construct \code{ShufArr}, and storing the secret permutation representation $\mathsecret{\vec{\pi}}$. We refer to this procedure as \prot{ShuffleMem}, which is the primary bottleneck during \sysname initialization. 
To enhance the efficiency, Section~\ref{subsec:shuffmem} provides a constant-round \prot{ShuffleMem} for the underlying $(2,3)$-secret shares.
(2) constructing the index map that translates the logical index $\mathsecret{i}$ into a physical index $p$ pointing to the target partition of \code{ShufArr}, where \code{Arr}$_i$ and \code{ShufArr}$_p$ refer to the same partition. 
The construction phase follows the same procedure as ORAM (see Section~\ref{subsec:goram-pre}).

\tinypara{Access the partition.} 
After the initialization, the MPC servers can jointly access \sysname as follows:
% Given a secret vertex or edge, the MPC servers can access the target partition through \sysname as follows:

(1) Given the target vertex $\mathsecret{v}$ (from the clients), the servers can compute the partition index $\mathsecret{\lceil \frac{v}{k} \rceil}$ and obtain the partition containing $bl$ secret edges by accessing \voram. This partition contains \emph{all} the direct neighbors of $\mathsecret{v}$.

(2) Given the target edge $(\mathsecret{v_s}, \mathsecret{v_d})$, the servers can compute the partition index $\mathsecret{\lceil \frac{v_s}{k} \rceil} * b + \mathsecret{\lceil \frac{v_d}{k}\rceil}$ and obtain the partition containing $l$ secret edges by accessing \eoram. This partition contains \emph{all} edges $(\mathsecret{v_s}, \mathsecret{v_d})$ of the global graph if the edges exist.

For each given query, the servers only need to access one partition and scan the single partition to obtain the result (see Section~\ref{sec:local_query}). Thus, the query processing complexity is sublinear.

\tinypara{Batched update. }
To update \sysname, each data provider can locally update the private graph $G_i$ and send the updated edge blocks to the computation servers.
The servers then update the corresponding edge blocks in the global graph $\mathsecret{\gb}$ and re-construct the secure index, similar to the construction procedure.

\subsection{Parallelization and Vectorization}\label{subsec:parall}

\sysname is a \emph{parallel-friendly} data structure. All stages in its lifecycle can be accelerated through parallel processing.

In the local process stage, each data provider can independently split the private graph $G_i$ into the 2d-partitioned format, during which the edge blocks can be processed in parallel.
During the global integration, the primary bottleneck, \ie the \emph{odd\_even\_merge\_sort} of $b^2$ edge blocks from $N^{+}$ secret graphs $\{\mathsecret{G_{u}.{\sf{Block}}}\}_{i=1}^{i=N^{+}}$, can be performed in at most $b^2$ tasks in parallel.
After obtaining the $b\times b \times l$ global graph partitions, we can split it into $p$ \emph{partition slices} for arbitrary $p \leq l$ because each edge block is a list of $l$ edges that can be processed in parallel.
Specifically, we split the $b \times b$ edge blocks by edges into $p$ $b \times b \times l^{(j)}$ partition slices, $l = \sum_{j=1}^{p}l^{(j)}$. Each slice contains all the edge blocks but fewer edges per block. We can then establish $p$ secure indices for each partition slice concurrently. 
For each query, the $p$ partition slices can be accessed and processed in parallel, and each slice can be processed as a single vector for better performance (Section~\ref{sec:local_query}). The query result can be obtained by merging the results of $p$ partition slices. 
\subsection{Optimization on ShuffleMem Step}\label{subsec:shuffmem}

The most expensive step during \sysname initialization is the \prot{ShuffleMem} procedure in secure index construction, which shuffles the array according to a random permutation $\pi$ and stores the secret permutation representation $\mathsecret{\vec{\pi}}$.
For an array with $n$ partitions, the original \prot{ShuffleMem} (\ie \emph{Waksman permutation network} adopted in \cite{zahur2016revisiting}) incurs $O(n \log n)$ communication and computation, which is expensive for large $n$.
To optimize this, \sysname designs a constant-round $O(n)$ \prot{ShuffleMem} to accelerate the initialization process by extending Araki \etal~\cite{araki2021secure} on $(2,3)$-secret shares.

\tinypara{The \prot{ShuffleMem} procedure. }
The computation servers begin with a secret shared array $\mathsecret{D} = \{\mathsecret{D_0}, \mathsecret{D_1}, \dots, \mathsecret{D_{n-1}}\}$ of $n$ partitions. 
At the end of the protocol, the computation servers output two secret shared arrays $\mathsecret{\widetilde{D}}$ and $\mathsecret{\vec{\pi}}$, where $\widetilde{D}$ is a permutation of $D$ under some random \emph{permutation} $\pi$ and $\mathsecret{\vec{\pi}}$ is the secret-shared \emph{permutation representation} of $\pi$.
The permutation $\pi$ is a bijection mapping from $D$ to itself that moves the $i$th partition $D_i$ to place $\pi(i)$. The permutation result $\widetilde{D} = \{\widetilde{D}_0, \widetilde{D}_1, \dots, \widetilde{D}_{n-1}\} = \pi(D)$ satisfies that $D_i = \widetilde{D}_{\pi(i)}, \forall i \in \{0, 1, \dots n-1\}$. 
The permutation representation $\vec{\pi}$ is an array of $n$ elements that explicitly records the location of each $D_i$ in $\widetilde{D}$ in its $i$th element $\vec{\pi}_i$.

\begin{Protocol}[t]
    % \footnotesize
    % \left
    % \raggedright
    \centering
    \resizebox{0.48\textwidth}{!}{
    \renewcommand{\arraystretch}{0.9}
    \footnotesize
    \begin{tabular}{|c c c c c c|}
    \hline
    \multicolumn{2}{|c}{$S1 (A, B)$} & \multicolumn{2}{c}{$S2 (B, C)$} & \multicolumn{2}{c|}{$S3 (C, A)$} \\ \hline

    \multicolumn{6}{|c|}{\textbf{0) Construct the shares of $L = [n]$.}} \\\hline 
    \multicolumn{2}{|c}{\highl{$L_A = Z_1 \xor L$}} &  \multicolumn{2}{c}{\highl{$L_B = Z_2$}} & \multicolumn{2}{c|}{\highl{$L_C = Z_3$}} \\ \hline 
    % \multicolumn{2}{|c}{} &  \multicolumn{2}{c}{$L_A = L_B \xor L_C \xor L$} & \multicolumn{2}{c|}{~} \\ \hline
    \multicolumn{1}{|l}{\highl{$\leftarrow L_A$}} & \multicolumn{1}{c}{~} & \multicolumn{1}{l}{\highl{$\leftarrow L_B$}}  & \multicolumn{1}{c}{~} & \multicolumn{2}{l|}{\highl{$\leftarrow L_C$}} \\ \hline
    
    \multicolumn{6}{|c|}{\textbf{1) Prepare the correlated randomness.}} \\\hline
    \multicolumn{2}{|c}{$Z_{12}$, \highl{$Z^{L}_{12}$}, $\tilde{B}$} &  \multicolumn{2}{c}{$Z_{12}$, \highl{$Z^{L}_{12}$}, $\tilde{B}$} & \multicolumn{2}{c|}{~} \\ 
    \multicolumn{2}{|c}{$\pi_{12}$ and \highl{$\pi^{-1}_{12}$}} &  \multicolumn{2}{c}{$\pi_{12}$ and \highl{$\pi^{-1}_{12}$}} & \multicolumn{2}{c|}{~} \\
    \multicolumn{2}{|c}{~} & \multicolumn{2}{c}{$Z_{23}$, \highl{$Z^{L}_{23}$, $\tilde{L_C}$}} & \multicolumn{2}{c|}{$Z_{23}$, \highl{$Z^{L}_{23}$, $\tilde{L_C}$}} \\
    \multicolumn{2}{|c}{~} & \multicolumn{2}{c}{$\pi_{23}$ and \highl{$\pi^{-1}_{23}$}} & \multicolumn{2}{c|}{$\pi_{23}$ and \highl{$\pi^{-1}_{23}$}} \\ 
    \multicolumn{2}{|c}{$Z_{31}$, \highl{$Z^{L}_{31}$}, $\tilde{A}$, \highl{$\tilde{L_A}$}} & \multicolumn{2}{c}{~} & \multicolumn{2}{c|}{$Z_{31}$, \highl{$Z^{L}_{31}$}, $\tilde{A}$, \highl{$\tilde{L_A}$}} \\
    \multicolumn{2}{|c}{$\pi_{31}$ and \highl{$\pi^{-1}_{31}$}} & \multicolumn{2}{c}{~} & \multicolumn{2}{c|}{$\pi_{31}$ and \highl{$\pi^{-1}_{31}$}} \\ \hline

    \multicolumn{6}{|c|}{\textbf{2) Main protocol: computation and communications}} \\\hline  
    
    % phase1
    \multicolumn{2}{|c}{$X_1 = \pi_{12}(A \oplus B \oplus Z_{12})$} & \multicolumn{2}{c}{$Y_1 = \pi_{12}(C \oplus Z_{12})$} & \multicolumn{2}{c|}{~} \\
    \multicolumn{2}{|c}{$X_2 = \pi_{31}(X_1 \oplus Z_{31})$} & \multicolumn{2}{c}{~} & \multicolumn{2}{c|}{~} \\
    \multicolumn{2}{|c}{~} & \multicolumn{2}{c}{\highl{$LY_1 = \pi^{-1}_{23}(L_B \oplus Z^{L}_{23})$}} & \multicolumn{2}{c|}{\highl{$LX_1 = \pi^{-1}_{23}(L_C \oplus L_A \oplus Z^{L}_{23})$}} \\
    \multicolumn{2}{|c}{~} & \multicolumn{2}{c}{~} & \multicolumn{2}{c|}{\highl{$LX_2 = \pi^{-1}_{31}(LX_1 \oplus Z^{L}_{31})$}} \\ \hline
    % communication
    \multicolumn{4}{|c}{$X_2 \leftrightarrow$ \highl{$LY_1$}}  &  \multicolumn{2}{l|}{$Y_1 \leftrightarrow$ \highl{$LX_2$}} \\ \hline
    
    % phase2
    \multicolumn{2}{|c}{~} & \multicolumn{2}{c}{~} & \multicolumn{2}{c|}{$Y_2 = \pi_{31}(Y_1 \oplus Z_{31})$} \\
    \multicolumn{2}{|c}{~} & \multicolumn{2}{c}{$X_3 = \pi_{23}(X_2 \oplus Z_{23})$} & \multicolumn{2}{c|}{$Y_3 = \pi_{23}(Y_2 \oplus Z_{23})$} \\ 
    \multicolumn{2}{|c}{~} & \multicolumn{2}{c}{$\tilde{C_1} = X_3 \oplus \tilde{B}$} & \multicolumn{2}{c|}{$\tilde{C_2} = Y_3 \oplus \tilde{A}$} \\
    \multicolumn{2}{|c}{\highl{$LY_2 = \pi_{31}^{-1}(LY_1 \oplus Z^{L}_{31})$}} & \multicolumn{2}{c}{~} & \multicolumn{2}{c|}{~} \\
    \multicolumn{2}{|c}{\highl{$LY_3 = \pi_{12}^{-1}(LY_2 \oplus Z^{L}_{12})$}} & \multicolumn{2}{c}{\highl{$LX_3 = \pi^{-1}_{12}(LX_2 \oplus Z^{L}_{12})$}} & \multicolumn{2}{c|}{~} \\
    \multicolumn{2}{|c}{\highl{$\tilde{L_{B_1}} = LY_3 \oplus \tilde{L_A}$}} & \multicolumn{2}{c}{\highl{$\tilde{L_{B_2}} = LX_3 \oplus \tilde{L_C}$}} & \multicolumn{2}{c|}{~} \\ \hline
    % communication
    \multicolumn{4}{|c}{\highl{$L_{B_1} \leftrightarrow L_{B_2}$}}  &  \multicolumn{2}{l|}{$\tilde{C_1} \leftrightarrow \tilde{C_2}$} \\ \hline
    
    % phase 3
    \multicolumn{2}{|c}{~} & \multicolumn{2}{c}{$\tilde{C} = \tilde{C_1} \oplus \tilde{C_2}$} & \multicolumn{2}{c|}{$\tilde{C} = \tilde{C_1} \oplus \tilde{C_2}$} \\
    \multicolumn{2}{|c}{\highl{$\tilde{L_B} = \tilde{L_{B_1}} \oplus \tilde{L_{B_2}}$}} & \multicolumn{2}{c}{\highl{$\tilde{L_B} = \tilde{L_{B_1}} \oplus \tilde{L_{B_2}}$}} & \multicolumn{2}{c|}{~} \\\hline
    
    \multicolumn{6}{|c|}{\textbf{3) Output}} \\\hline 
    \multicolumn{2}{|c}{$\tilde{A}, \tilde{B}$, \highl{$\tilde{L_A}$, $\tilde{L_B}$}} & \multicolumn{2}{c}{$\tilde{B}, \tilde{C}$, \highl{$\tilde{L_B}$, $\tilde{L_C}$}} & \multicolumn{2}{c|}{$\tilde{C}, \tilde{A}$, \highl{$\tilde{L_C}$, $\tilde{L_A}$}} \\\hline 

    \end{tabular}
    }
    \vspace{-.05in}
    \caption{\prot{ShuffleMem} Build Protocol $\Pi_{\text{ShufMem}}$ {\small{- \highl{Lightgray} operations are our extensions of Araki \etal~\cite{araki2021secure} to compute $\mathsecret{\vec{\pi}}$}.}}\label{prot:shuffleMem}
    \vspace{-.13in}
\end{Protocol}

\tinypara{Key idea of constant-round construction.} 
We design the constant-round \prot{ShuffleMem} by extending the constant-round shuffle protocol of Araki \etal~\cite{araki2021secure}, which computes $\mathsecret{\widetilde{D}}$ in $O(n)$ complexity and $O(1)$ communication rounds using $(2,3)$-secret sharing.
The key idea of our protocol is to compute $\mathsecret{\vec{\pi}}$ simultaneously by leveraging the properties of permutations:

\begin{enumerate}[itemindent=-0.1cm, labelsep=0.08cm, label=$\ast$, itemsep=0.01cm]
    \item Permutations are composable, \ie $\pi_1 \circ \pi_2$ is also a permutation such that $(\pi_1 \circ \pi_2)(x) = \pi_1(\pi_2(x))$ given array $x$.
    \item Permutations are inversible, for each permutation $\pi$, there exists $\pi^{-1}$ such that $(\pi^{-1} \circ \pi)(x) \equiv x$.
    \item The permutation representation $\vec{\pi} = \pi^{-1}(L)$, where $L = \{0,1,\dots,n-1\}$, $n$ is the size of $x$. 
\end{enumerate}

Specifically, Araki \etal~\cite{araki2021secure} implement the random shuffle by letting the computation servers collaboratively shuffle the data using three random permutations $\pi_{12}$, $\pi_{23}$ and $\pi_{31}$, \ie $\mathsecret{\widetilde{D}} = \pi_{23} \circ \pi_{31} \circ \pi_{12}(\mathsecret{D}) = \pi({\mathsecret{D}})$. The permutation $\pi_{ij}$ is only known to servers $S_i$ and $S_j$. Because each computation server only knows two out of the three random permutations, the overall permutation $\pi = \pi_{23} \circ \pi_{31} \circ \pi_{12}$ remains random for each computation server. 
Using the same collaborative shuffle procedure, we can compute the secret permutation representation $\mathsecret{\vec{\pi}}$ simultaneously by shuffling the ranging array $L = \{0,1,\dots,n-1\}$ using the inverse permutations \ie $\mathsecret{\vec{\pi}} = \pi^{-1}(\mathsecret{L}) = \pi^{-1}_{12} \circ \pi^{-1}_{31} \circ \pi^{-1}_{23}(\mathsecret{L})$.

\tinypara{\prot{ShuffleMem} construction. } Protocol~\ref{prot:shuffleMem} shows the \prot{ShuffleMem} construction.
Each pair of computation servers $S_i$ and $S_j$ share a common random seed $s_{i,j}$ beforehand.
As inputs to this protocol, each computation server holds two out of the three shares $A, B, C$, satisfying that the input $D \equiv A \xor B \xor C$. 
Also, the computation servers construct the shares of the ranging array $L \equiv L_A \xor L_B \xor L_C$, which requires one round of communication. 
Specifically, $S_1, S_2$ and $S_3$ at first construct an array of secret shares on zeros, \ie $Z_1 \xor Z_2 \xor Z_3 \equiv \vec{0}$, $|\vec{0}| = n$, which requires no interactions using~\cite{mohassel2018aby3}. Each $Z_i$ is uniformly random and is only known to $S_i$. 
$S_1$ locally computes $L_A = Z_1 \xor L$ and each server sends its share to the previous server to obtain the secret shares of $L$ (step 0).
The first step of Protocol~\ref{prot:shuffleMem} is to set up the correlated randomness using the pairwise random seed. 
Also, each pair of $S_i$ and $S_j$ generates a random permutation $\pi_{i, j}$ and the inverse permutation $\pi^{-1}_{i, j}$. 

The computation servers then begin the main protocol, during which there are two invariants held: 
1) $X_i \xor Y_i$ is a permutation of $D$ and
2) $LX_i \xor LY_i$ is an inverse permutation of $L$. 
For examples, $X_1 \xor Y_1 = \pi_{12}(A \xor B \xor Z_{12}) \xor \pi_{12}(C \xor Z_{12}) = \pi_{12}(A \xor B \xor C) = \pi_{12}(D)$, $LY_1 \xor LX_1 = \pi^{-1}_{23}(L_B \xor Z^{L}_{23}) \xor \pi^{-1}_{23}(L_C \xor L_A \xor Z^{L}_{23}) = \pi^{-1}_{23}(L)$. That is, during the main protocol, the servers sequentially compute $X_1 \xor Y_1 = \pi_{12}(D)$, $X_2 \xor Y_2 = (\pi_{31} \circ \pi_{12})(D)$ and $X_3 \xor Y_3 = (\pi_{23} \circ \pi_{31} \circ \pi_{12})(D)$, which constitutes the final shares of $\pi(D)$, $\pi = \pi_{23} \circ \pi_{31} \circ \pi_{12}$. 
The permutation representation $\vec{\pi} = \pi^{-1}(L) = (\pi^{-1}_{12} \circ \pi^{-1}_{31} \circ \pi^{-1}_{23})(L)$ is constructed similarly in the reverse order. 

% \hz{It's very challenging for readers like me who do not have a solid background in MPC to understand the details in this protocol. I don't know some of the key concepts and notations. For example, I have no idea what a \emph{zero share} $Z$ is, and if $L$ represents a set and $\xor$ represents XOR, what does it mean to have a zero share XOR a set? The questions may seem dumb, but I don't know whether this is common for DB reviewers, or it's just me. Maybe adding a small section in the background to introduce the key concepts in MPC would be a good idea?}

\tinypara{Correctness. } From the two invariants, it is straightforward to see the correctness of \prot{ShuffleMem} Protocol~\ref{prot:shuffleMem}. Because the final shares satisfy that $\tilde{A} \xor \tilde{B} \xor \tilde{C} = \tilde{X_3} \xor \tilde{Y_3} = \pi(D)$, and $\tilde{L_A} \xor \tilde{L_B} \xor \tilde{L_C} = \tilde{LX_3} \xor \tilde{LY_3} = \pi^{-1}(L)$, the correctness is guaranteed.

% \psec{
% \tinypara{Security. } The semi-honest security of Protocol~\ref{prot:shuffleMem} can be proved through the \emph{real-ideal} paradigm~\cite{canetti2000security}, similar to the proof of the original shuffle protocol~\cite{araki2021secure}. Briefly, security is guaranteed because \emph{all} the messages sent to each server $S_i$ are uniformly random in the view of $S_i$. This is achieved by masking all messages that $S_i$ receives using at least one randomness that $S_i$ does not know. E.g., $X_2$ sent to $S_2$ is masked using $Z_{31}$, which is random to $S_2$, thereby making $X_2$ random.
% Therefore, these parties can simulate all the received messages by random sampling. 
% In fact, it is intuitive that Protocol~\ref{prot:shuffleMem} preserves the same security as \cite{araki2021secure}, since its extension to apply the inverse permutation on the secret shares of the ranging array $L$ follows the same permutation method in \cite{araki2021secure}.
% } 

\tinypara{Security. } Protocol~\ref{prot:shuffleMem} satisfies Theorem~\ref{thm:sec-shufflemem}. We provide a sketch proof here and the complete proof is shown in Appendix~\ref{app:shuffle-sec}.

\begin{theorem}\label{thm:sec-shufflemem}
    Protocol~\ref{prot:shuffleMem} securely implements the \prot{ShuffleMem} procedure against any semi-honest adversary controlling at most one computation server.
\end{theorem}

\begin{proof}(sketch)
    We prove Theorem~\ref{thm:sec-shufflemem} with the \emph{real-ideal} paradigm~\cite{canetti2000security}. 
    Let $\mathcal{A}$ denotes the real-world adversary and $\mathcal{S}$ denotes the simulator, which simulates the view of $\mathcal{A}$ in the ideal world. Protocol~\ref{prot:shuffleMem} is secure if for all $\mathcal{A}$, there exists a simulator $\mathcal{S}$, such that for all inputs and for all corrupted party $S_i, i \in [3]$, 

    \begin{equation}\label{eq:sec-view}
        \text{View}(\mathcal{A}) \equiv \text{View}(\mathcal{S})
    \end{equation}

    For each possible corrupted $S_i$, all the messages it receives are uniformly random in its view. This is achieved by masking all the messages $S_i$ receives with at least one randomness that $S_i$ does not know. 
    Therefore, we can construct the simulator $\mathcal{S}$ by randomly sampling all the messages that $\mathcal{A}$ receives. The $\text{View}(\mathcal{S})$ is uniformly random and therefore indistinguishable from $\text{View}(\mathcal{A})$.
\end{proof}

\begin{algorithm}[t]
	\small
        % \SetKwInOut{Global}{\textbf{Global}}
	\SetKwInOut{Input}{\textbf{Inputs}}
	\SetKwInOut{Output}{\textbf{Output}}
        % \SetKwInOut{UDF}{\textbf{UDFs}} 
	% \caption{\small \code{EdgeAnalyze}}\label{algo:edge_exist}
        \caption{\small \code{EdgeExist} (MPC servers compute)}\label{algo:edge_exist}
        % \Global{Constructed \code{GORAM}.}
	\Input{Target edge $(\mathsecret{v_s}, \mathsecret{v_d})$.}
	\Output{$\mathsecret{{\sf{flag}}}$ indicating whether the target edge exist in global $G$. }
	% \BlankLine
        % \comm{\cfont{The MPC servers compute the following. }}
        Compute the secret partition ID $\mathsecret{i} = \mathsecret{\lceil \frac{v_s}{k} \rceil * b + \lceil \frac{v_d}{k}\rceil}$\;
        % \algorithmiccomment{Partition extraction. }
        % \BlankLine
        Fetch the target edge partition $\mathsecret{B} \leftarrow \code{EORAM}.{\sf{access}}(\mathsecret{i})$, where $\mathsecret{B}$ contains $l$ $\sf{source\_nodes}$ and $\sf{dest\_nodes}$\;
        % \BlankLine
        \comm{\cfont{Vectorized edges comparisons. }}
        Construct $\mathsecret{\vec{v_s}}$ and $\mathsecret{\vec{v_d}}$ by expanding  $\mathsecret{v_s}$ and $\mathsecret{v_d}$ $l$ times\;
        Compute $\mathsecret{{\sf{mask}_{s}}} \leftarrow {\sf{EQ}}(\mathsecret{\vec{v_s}}, \mathsecret{B}.{\sf{source\_nodes}})$ \;
        Compute $\mathsecret{{\sf{mask}_{d}}} \leftarrow {\sf{EQ}}(\mathsecret{\vec{v_d}}, \mathsecret{B}.{\sf{dest\_nodes}})$ \;
        % Compute $\mathsecret{\sf{mask}_{edg}} \leftarrow {\sf{AND}(\mathsecret{{\sf{mask}_{s}}}, \mathsecret{{\sf{mask}_{d}}})}$ \;
        Compute $\mathsecret{\sf{mask}} \leftarrow {\sf{AND}(\mathsecret{{\sf{mask}_{s}}}, \mathsecret{{\sf{mask}_{d}}})}$ \;
        % \tcc{Filtering UDF specific edges.  }
        % Compute $\mathsecret{{\sf{mask}_{cond}}} \leftarrow {\sf{cond}}(\mathsecret{B}.{\sf{property}})$ \;
        % Compute $\mathsecret{\sf{mask}} \leftarrow {\sf{AND}(\mathsecret{{\sf{mask}}}, \mathsecret{{\sf{mask}}_{cond}})}$ \;
        % \tcc{Aggregating the result from filtered properties. }
        % $\mathsecret{\sf{property}_{mask}} \leftarrow {\sf{AND}(\mathsecret{B}.{\sf{property}}, \mathsecret{\sf{mask}})}$ \;
        % $\mathsecret{\sf{res}} = {\sf{agg}(\mathsecret{\sf{property}_{mask}})}$\;
        \comm{\cfont{Aggregating the result through {\sf{OR}}. }}
        \While{$\sf{len}(\mathsecret{\sf{mask}}) > 1$}{
            Pad $\mathsecret{0}$ to $\mathsecret{\sf{mask}}$ to be even \;
            Split $\mathsecret{\sf{mask}}$ half-by-half to $\mathsecret{\sf{mask}}_l$ and $\mathsecret{\sf{mask}}_r$\;
            Aggregate $\mathsecret{\sf{mask}} \leftarrow {\sf{OR}}(\mathsecret{\sf{mask}}_l, \mathsecret{\sf{mask}}_r)$ \;
        }
        Compute $\mathsecret{{\sf{flag}}} = \mathsecret{\sf{mask}}$\;
        \Return{${\mathsecret{\sf{flag}}}$} \text{to the client.}
\end{algorithm}

\begin{algorithm}[t]
	\small
	\SetKwInOut{Input}{\textbf{Inputs}}
	\SetKwInOut{Output}{\textbf{Output}}
        % \SetKwInOut{UDF}{\textbf{UDFs}} 
	% \caption{\small \code{NeighborsAnalyze}}\label{algo:neighbors_count}
        \caption{\small \code{NeighborsCount} (MPC servers compute)}\label{algo:neighbors_count}
        % \Global{Constructed \code{GORAM}.}
	\Input{Target vertex $\mathsecret{v}$.}
        % \UDF{{\sf{cond}}($\mathsecret{{\sf{property}}}$) $\rightarrow \mathsecret{{\sf{mask}_{cond}}}$: conditional mask of the target edges.\\
        % {\sf{agg}({$\mathsecret{\sf{property}}$}) $\rightarrow \mathsecret{\sf{res}}$}: aggregation function\\ for statistic result.}
	% \Output{$\mathsecret{{\sf{flag}}}$ indicating whether the target edge exist in global $G$ or not. }
	\Output{$\mathsecret{{\sf{num}}}^{A}$, the number of $v$'s outing neighbors. }
	% \BlankLine
        % \comm{\cfont{The MPC servers compute the following. }}
        % \BlankLine
        Compute the secret partition ID $\mathsecret{i} = \mathsecret{\lceil \frac{v}{k} \rceil}$\;
        Fetch the target edge partition $\mathsecret{B} \leftarrow \code{VORAM}.{\sf{access}}(\mathsecret{i})$, where $\mathsecret{B}$ contains $(bl)$ $\sf{source\_nodes}$ and $\sf{dest\_nodes}$\;
        % \BlankLine
        \comm{\cfont{Filtering real neighbors using vectorization. }}
        Construct $\mathsecret{\vec{v}}$ by expanding  $\mathsecret{v}$ $bl$ times\;
        Compute $\mathsecret{{\sf{mask}}} \leftarrow {\sf{EQ}}(\mathsecret{\vec{v}}, \mathsecret{B}.{\sf{source\_nodes}})$ \;
        Obtain the arith shares $\mathsecret{\prot{mask}}^{A} \leftarrow \prot{B2A}{(\mathsecret{{\sf{mask}}})}$ \;
        % \tcc{Filtering UDF specific properties. }
        % Compute $\mathsecret{{\sf{mask}_{cond}}} \leftarrow {\sf{cond}}(\mathsecret{B}.{\sf{property}})$ \;
        % Compute $\mathsecret{{\sf{mask}}} \leftarrow {\sf{AND}(\mathsecret{{\sf{mask}_{neigh}}}, \mathsecret{{\sf{mask}}_{cond}})}$ \;
        \comm{\cfont{Counting real neighbor masks. }}
        % $\mathsecret{{\sf{property}_{mask}}} \leftarrow {\sf{AND}(\mathsecret{B}.{\sf{property}}, \mathsecret{{\sf{mask}}})}$\;
        % $\mathsecret{{\sf{res}}} \leftarrow {\sf{agg}(\mathsecret{{\sf{property}_{mask}}})}$\;
        Compute $\mathsecret{{\sf{num}}}^{A} \leftarrow {\sf{SUM}}(\mathsecret{\prot{mask}}^{A})$\;
        \Return{$\mathsecret{{\sf{num}}}^{A}$} to the client.
\end{algorithm}

\begin{table*}[t]
% \small
\caption{Complexity Summarization (see Appendix~\ref{app:complexity} for detailed derivation. )}\label{tab:complexity}
\vspace{-.1in}
\resizebox{0.99\textwidth}{!}{
\begin{threeparttable}
\renewcommand{\arraystretch}{1.7}

\begin{tabular}{cc|cc|cc|cccccc}
\midrule[1.1pt]
\multicolumn{2}{c|}{\multirow{3}{*}{\normalsize{Data Structures}}} &
  \multicolumn{2}{c|}{\multirow{2}{*}{\normalsize{Initialization}}} &
  \multicolumn{2}{c|}{\multirow{2}{*}{\normalsize{Partition Access}}} &
  \multicolumn{6}{c}{\normalsize{Partition Processing for Basic Queries}} \\ \cline{7-12} 
\multicolumn{2}{c|}{} &
  \multicolumn{2}{c|}{} &
  \multicolumn{2}{c|}{} &
  \multicolumn{2}{c|}{\code{EdgeExist}} &
  \multicolumn{2}{c|}{\code{NeighborsCount}} &
  \multicolumn{2}{c}{\code{NeighborsGet}} \\ \cline{3-12} 
\multicolumn{2}{c|}{} &
  \multicolumn{1}{c|}{Comp} &
  Round &
  \multicolumn{1}{c|}{Comp} &
  Round &
  \multicolumn{1}{c|}{Comp} &
  \multicolumn{1}{c|}{Round} &
  \multicolumn{1}{c|}{Comp} &
  \multicolumn{1}{c|}{Round} &
  \multicolumn{1}{c|}{Comp} &
  Round \\ \hline \hline
\multicolumn{1}{c|}{\multirow{2}{*}{\mat}} &
  adj-\voram &
  \multicolumn{1}{c|}{\multirow{2}{*}{$O(N{|V|}^2)$}} &
  3 +2$\log_P(\frac{|V|}{T})$ &
  \multicolumn{1}{c|}{$O(PT\log_P(\frac{|V|}{T}))$} &
  $O(\log_P(\frac{|V|}{T}))$ &
  \multicolumn{1}{c|}{\multirow{2}{*}{$O(1)$}} &
  \multicolumn{1}{c|}{\multirow{2}{*}{$O(1)$}} &
  \multicolumn{1}{c|}{\multirow{2}{*}{$O(|V|)$}} &
  \multicolumn{1}{c|}{\multirow{2}{*}{$O(1)$}} &
  \multicolumn{1}{c|}{\multirow{2}{*}{$O(|V|)$}} &
  \multirow{2}{*}{$O(1)$} \\ \cline{2-2} \cline{4-6}
\multicolumn{1}{c|}{} &
  adj-\eoram &
  \multicolumn{1}{c|}{} &
  3 +2$\log_P(\frac{|V|^2}{T})$ &
  \multicolumn{1}{c|}{$O(PT\log_P(\frac{|V|^2}{T}))$} &
  $O(\log_P(\frac{|V|^2}{T}))$ &
  \multicolumn{1}{c|}{} &
  \multicolumn{1}{c|}{} &
  \multicolumn{1}{c|}{} &
  \multicolumn{1}{c|}{} &
  \multicolumn{1}{c|}{} &
   \\ \hline 
\multicolumn{2}{c|}{\elist} &
  \multicolumn{1}{c|}{$O(|E|\log(|E|)\log(N))$} &
  $\log(|E|)\log(N)$ &
  \multicolumn{1}{c|}{$O(1)$} &
  NA &
  \multicolumn{1}{c|}{$O(|E|)$} &
  \multicolumn{1}{c|}{$O(\log(|E|))$} &
  \multicolumn{1}{c|}{$O(|E|)$} &
  \multicolumn{1}{c|}{$O(1)$} &
  \multicolumn{1}{c|}{$O(|E|)$} &
  $O(1)$ \\ \hline
\multicolumn{1}{c|}{\multirow{2}{*}{\sysname}} &
  \voram &
  \multicolumn{1}{c|}{\multirow{2}{*}{\makecell{$O(b^2l\log(l)\log(N^{+}))$}}} &
  3 + $\log(l)\log(N^{+})$+ 2$\log_P(\frac{b}{T})$ &
  \multicolumn{1}{c|}{$O(PT\log_P(\frac{b}{T}))$} &
  $O(\log_P(\frac{b}{T}))$ &
  \multicolumn{1}{c|}{\multirow{2}{*}{$O(l)$}} &
  \multicolumn{1}{c|}{\multirow{2}{*}{$O(\log(l))$}} &
  \multicolumn{1}{c|}{\multirow{2}{*}{$O(bl)$}} &
  \multicolumn{1}{c|}{\multirow{2}{*}{$O(1)$}} &
  \multicolumn{1}{c|}{\multirow{2}{*}{$O(bl)$}} &
  \multirow{2}{*}{$O(1)$} \\ \cline{2-2} \cline{4-6}
\multicolumn{1}{c|}{} &
  \eoram &
  \multicolumn{1}{c|}{} &
  3 + $\log(l)\log(N^{+})$+ 2$\log_P(\frac{b^2}{T})$ &
  \multicolumn{1}{c|}{$O(PT\log_P(\frac{b^2}{T}))$} &
  $O(\log_P(\frac{b^2}{T}))$ &
  \multicolumn{1}{c|}{} &
  \multicolumn{1}{c|}{} &
  \multicolumn{1}{c|}{} &
  \multicolumn{1}{c|}{} &
  \multicolumn{1}{c|}{} &
   \\ \hline
\end{tabular}

\begin{tablenotes}
    \item $P$ and $T$ denote the pack and stash size of ORAM. $b, l$ are the configuration parameters of 2d-partition, where $b = \frac{|V|}{k} = \frac{|E|}{|V|} = D$ is the graph density. The Round complexities with the $O(\cdot)$ notation is in the unit of secure operations like \prot{EQ}, and the Partition Access complexities are the averaged complexity of successive $T$ queries. $N \geq 1$ is the number of data providers. $N^{+}$ is the number of all the submitted partitions, $N^{+} \geq N$. 
    \end{tablenotes}
\end{threeparttable}
}
% \vspace{-.13in}
\end{table*}

\begin{algorithm}[t]
	\small
        \SetKwInOut{Global}{\textbf{Global}}
	\SetKwInOut{Input}{\textbf{Inputs}}
        % \SetKwInOut{UDF}{\textbf{UDFs}} 
	\SetKwInOut{Output}{\textbf{Output}}
	\caption{\small \code{NeighborsGet} (MPC servers compute)}\label{algo:neighbors_get}
	\Input{Target vertex $\mathsecret{v}$.}
        % \UDF{{\sf{cond}}($\mathsecret{{\sf{property}}}$) $\rightarrow \mathsecret{{\sf{mask}_{cond}}}$: conditional mask of the target edges. }
	\Output{$\mathsecret{{\sf{neighbors}}}$, the unique outing neighbor's IDs of $\mathsecret{v}$. }
	% \BlankLine
        % \comm{\cfont{MPC servers compute the following. }}
        % \tcc{Partition access. }
        % \BlankLine
        Compute the secret partition ID $\mathsecret{i} = \mathsecret{\lceil \frac{v}{k} \rceil}$\;
        Fetch the target edge blocks $\mathsecret{B} \leftarrow \code{VORAM}.{\sf{access}}(\mathsecret{i})$, where $\mathsecret{B}$ contains $(bl)$ $\sf{source\_nodes}$ and $\sf{dest\_nodes}$\;
        % \BlankLine
        % \tcc{Parallel and batched process. }
        % \BlankLine
        \comm{\cfont{1) Filtering real neighbors.}}
        Construct $\mathsecret{\vec{v}}$ by expanding $\mathsecret{v}$ $bl$ times\;
        % Compute $\mathsecret{{\sf{mask}_{neigh}}} \leftarrow {\sf{EQ}}(\mathsecret{\vec{v_s}}, \mathsecret{B}.{\sf{source\_nodes}})$ \;
        Compute $\mathsecret{{\sf{mask}}} \leftarrow {\sf{EQ}}(\mathsecret{\vec{v_s}}, \mathsecret{B}.{\sf{source\_nodes}})$ \;
        % \tcc{2) Compute UDF specific mask.}
        % Compute $\mathsecret{{\sf{mask}_{cond}}} \leftarrow {\sf{cond}}(\mathsecret{B}.{\sf{property}})$ \;
        % \comm{\cfont{2) Obtaining the neighbors and masking the others out.}}
        % Compute $\mathsecret{{\sf{mask}}} \leftarrow {\sf{AND}(\mathsecret{{\sf{mask}_{neigh}}}, \mathsecret{{\sf{mask}}_{cond}})}$ \;
        Compute $\mathsecret{{\sf{candidate}}} \leftarrow {\sf{MUL}}(\mathsecret{{\sf{mask}}}, \mathsecret{B}.{\sf{dest\_nodes}})$ \;
        \comm{\cfont{2) De-duplicating neighbors.}}
        $\mathsecret{{\sf{same\_mask}}} \leftarrow {\sf{NEQ}}(\mathsecret{{\sf{candidate}}}_{[1:]}, \mathsecret{{\sf{candidate}}}_{[:-1]})$\;
        Compute $\mathsecret{{\sf{same\_mask}}}.{\sf{append}(\mathsecret{1})}$\;
        % $\mathsecret{{\sf{same\_mask}}} \leftarrow {\sf{AND}}(\mathsecret{{\sf{same\_mask}}}, \mathsecret{{\sf{mask}}})$\;
        Compute $\mathsecret{\sf{neighbors}} \leftarrow {\sf{MUL}}(\mathsecret{{\sf{same\_mask}}}, \mathsecret{{\sf{candidate}}})$\;
        % \comm{\cfont{3) Shuffling the neighbors.}}
        Compute $\mathsecret{\sf{neighbors}} \leftarrow {\sf{SHUFFLE}}(\mathsecret{\sf{neighbors}})$\;
        \Return{$\mathsecret{\sf{neighbors}}$} to the client.
\end{algorithm}

\section{Querying Graphs through GORAM}\label{sec:local_query}

We provide five ego-centric query examples through \sysname, covering all queries listed in LinkBench~\cite{linkbench2013}. The other queries can be similarly achieved.
For each query, the client submits the secret query key to the servers. The servers process the queries and return the secret shares of the results to the client. Only the client can reconstruct the plaintext final result.

\subsection{Basic Queries}\label{subsec:basic_query}

\tinypara{EdgeExist} Algorithm~\ref{algo:edge_exist} is a basic query that checks whether an edge $(v_s, v_d)$ exists. We first access \eoram using secret index $\mathsecret{\lceil \frac{v_s}{k} \rceil}*b + \mathsecret{\lceil \frac{v_d}{k} \rceil}$. 
Then, we compare all the edges in the partition using vectorization. The comparison result is a secret $\mathsecret{\sf{mask}}$ indicating which edge is equivalent to the given edge. We obtain the result by aggregating $\mathsecret{\sf{mask}}$ using the \prot{OR} operation.

\tinypara{NeighborsCount} Algorithm~\ref{algo:neighbors_count} counts the number of target vertex $v$'s outing neighbors. Because the query is about vertex $v$, we refer to \voram for the partition containing all its outing neighbors using secret index \secret{\lceil \frac{v}{k} \rceil}. We obtain the result by comparing all starting vertices to $v$ and summing up the comparison result.
For efficiency, we transform $\mathsecret{\prot{mask}}$ to arithmetic shares, \ie  $\mathsecret{\prot{mask}}^A$, which enable summation without communications.

\tinypara{NeighborsGet} Algorithm~\ref{algo:neighbors_get} extracts all the 1-hop outing neighbors of the target vertex $v$ while maintaining the number of edges between each neighbor and $v$ private.
The first 3 lines access the target partition and compare all the starting vertices with $v$ to construct $\mathsecret{\prot{mask}}$, indicating the edges started from $v$. 
Then, we multiply the $\mathsecret{\prot{mask}}$ and the destination vertices to obtain $\mathsecret{\prot{candidate}}$, where each element is $\mathsecret{0}$ or $\mathsecret{u}$ if $u$ is an outing neighbor of $v$. Note that the number of $\mathsecret{u}$ implies the number of edges between $u$ and $v$, we then de-duplicate $\mathsecret{\prot{candidate}}$ in lines 5-8 to mask out this information.
Because each partition is sorted by key $v || u$ in the construction stage (see Section~\ref{subsec:cons_and_access}), all the same outing neighbors in $\mathsecret{\prot{candidate}}$ are located successively as a group. 
We apply a shifted \prot{NEQ} on $\mathsecret{\prot{candidate}}$ to compute $\mathsecret{\prot{same\_mask}}$, where only the last neighbor in each group is $\mathsecret{1}$, while the rest are $\mathsecret{0}$. 
By multiplying $\mathsecret{\prot{same\_mask}}$ and $\mathsecret{\prot{candidate}}$, we mask out the duplicate neighbors as $\mathsecret{0}$. Note that the gap between two successive neighbors $u_i, u_{i+1}$ still implies how many $u_{i+1}$ exist, and therefore, we apply \prot{SHUFFLE} to permute this location-implied information.

\subsection{Complex queries}
We provide two complex queries by extending the above queries.

\tinypara{Cycle-identification.} Identifying whether the bank transactions across multiple suspicious accounts form a cycle is an effective way for money laundering detection~\cite{neo4financial, qiu2018real}. Cycle identification can be achieved by submitting a series of \code{EdgeExist} queries. For example, given three vertices $v_1, v_2$ and $v_3$, by submitting \code{EdgeExist} queries on edges $(v_1, v_2), (v_2, v_3), (v_3, v_1)$ and their reverse edges, the client can detect whether a cycle exists among the three vertices. 
% Similarly, the clients can submit more \code{EdgeExist} queries for more vertices.

\tinypara{Neighbors-filtering} queries are one of the most common queries on graphs with attributes, \ie each edge has attributes like creation timestamp and transaction amounts. We can implement these queries by extending the basic queries with filters.
For instance, to perform \emph{association range queries}~\cite{linkbench2013} that count the outing edges created after a provided timestamp, we can extend \code{NeighborsCount} with an extra comparison to compute whether the creation timestamp is greater than the given timestamp before counting the result.
Specifically, we compute $\mathsecret{\prot{t\_mask}} \leftarrow \prot{GT}(\mathsecret{\text{timestamp field}}, \mathsecret{\text{given threshold}})$ and update the neighbors mask in the 3rd line of Algorithm~\ref{algo:neighbors_count} to $\mathsecret{\prot{mask}} = \prot{AND}(\mathsecret{\prot{mask}}, \mathsecret{\prot{t\_mask}})$. Then we obtain the number of outing edges created after the given timestamp.

% By summing up the $\mathsecret{\prot{mask}}$ in the end,we obtain the number of outing edges created after the given timestamp.

\section{Security and Complexity Analysis}\label{sec:analysis}

\subsection{Security Analysis}\label{subsec:security}

We illustrate the security guarantees of \sysname by showing that no involved party (data providers, clients, and computation servers) can learn anything about the private graphs and the query keys.

(1) Data providers learn nothing about the others' private graphs or the query keys because they only submit secret-shared graph partitions to the servers.

(2) Clients learn nothing about the data providers' graphs or the other keys because they only submit secret-shared keys to the servers and receive results obtained from the global graph. 
% Therefore, they learn nothing about the private graph of each data provider.

(3) During \sysname initialization and query process, the computation servers only execute MPC protocols on $N^{+}$ indistinguishable $b \times b \times \bar{l}$ secret-shared partitioned graph matrices and secret-shared query keys. These protocols are executed in the standard \emph{modular composition} manner~\cite{canetti2000security}, which ensures that the overall computation performed by the servers inherits the same security guarantees of the underlying protocols. This includes the ABY3 protocols~\cite{mohassel2018aby3} and the proposed \prot{ShuffleMem} protocol, whose security is proved in Section~\ref{subsec:shuffmem}.
Consequently, the only potential source of private information for the servers is the sizes of the input matrices, \ie $b$, $\bar{l}$ and the total matrices number $N^{+}$. Among these, $\bar{l}$ is a public parameter, and $b = \lceil \frac{|V|}{k} \rceil$ is derived from two public parameters, $|V|$ and $k$. The number of the matrices, $N^{+}$, only reveals the partition size of the global graph and cannot be traced back to any specific data provider because of the anonymous authentication property provided by ring signature.
Therefore, both the private graphs and the query keys are protected throughout the lifecycle of \sysname.

\begin{figure*}[t]
	\centering
	\includegraphics[width=0.9\linewidth]{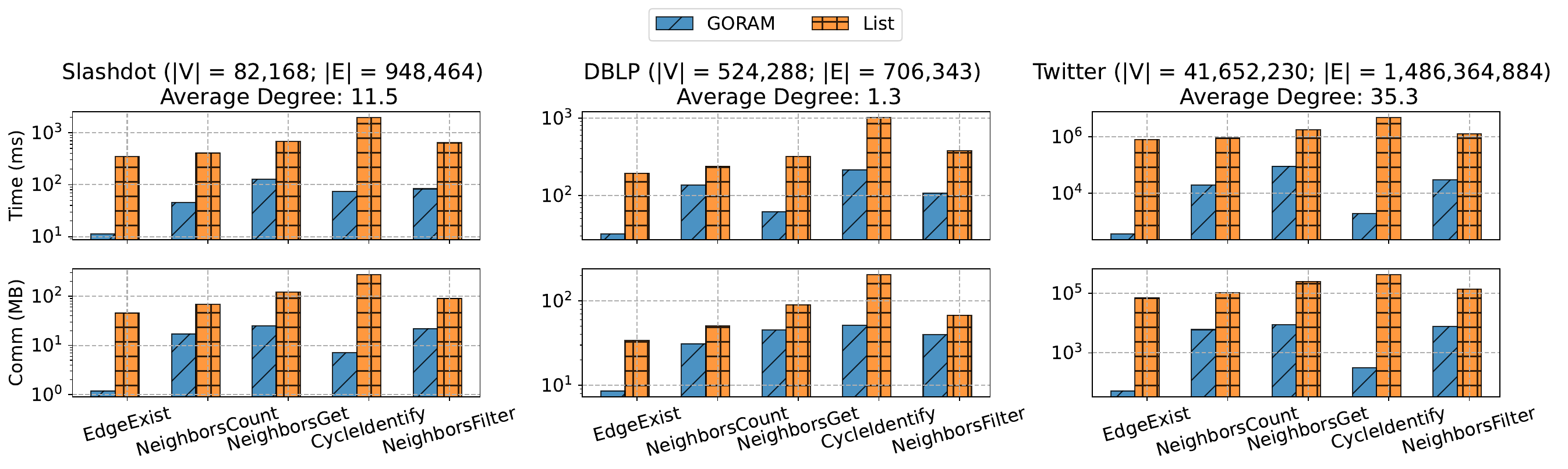}
	\vspace{-.1in}
	\caption{Queries on Real-world Graphs {\footnotesize * the y-axes are in log-scale. Queries on Twitter are in 16 threads and the others are single-threaded.}}
	\label{fig:perf_realworld}
\end{figure*}

\begin{figure*}[t]
	\centering
	\includegraphics[width=0.9\linewidth]{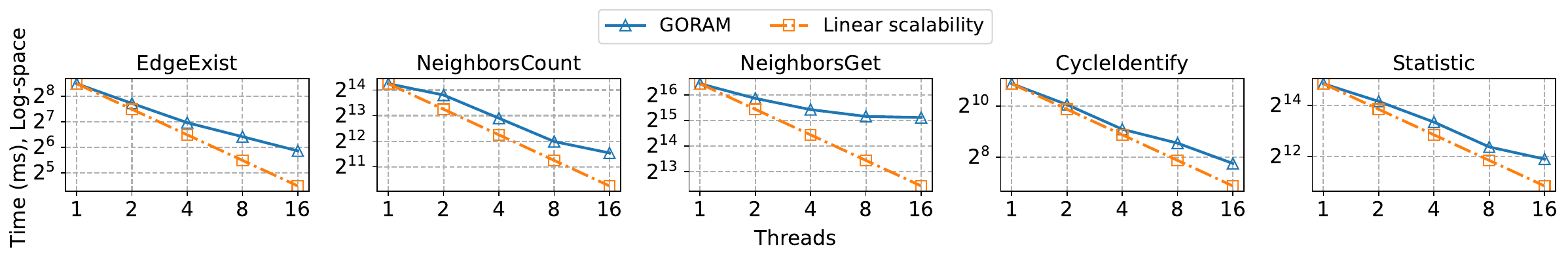}
	\vspace{-.1in}
	\caption{Parallelism on Twitter~\cite{twitter} (Queries)}
	\label{fig:parallel}
        \vspace{-.1in}
\end{figure*}

\begin{figure}[t]
    \centering
    \includegraphics[width=0.45\textwidth]{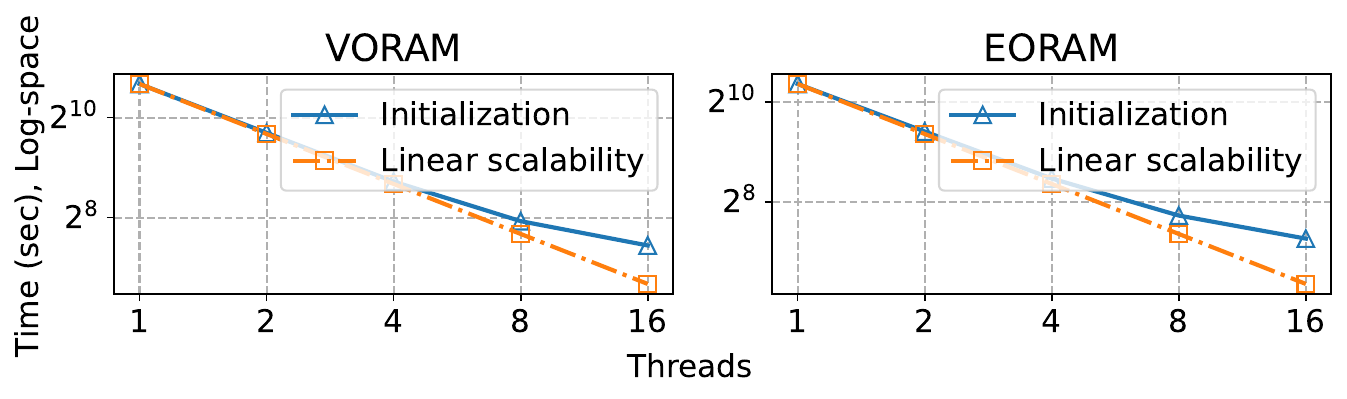}
    \vspace{-.15in}
    \caption{Parallelism on Twitter~\cite{twitter} (Initialization)}
    \label{fig:parallel_goram}
    \vspace{-.1in}
\end{figure}

\subsection{Complexities Analysis}\label{sec:complexity}

\tinypara{Complexity summarization. } Table~\ref{tab:complexity} summarizes the complexities of initialization, partition access, and basic query processing using \sysname and the strawman solutions. 
The complexities are directly derived from the algorithms of their respective sections. 
Specifically, \sysname's initialization is bottlenecked by the $O(b^2l\log(l)\log(N^{+}))$ merge sort on $N^{+}$ partitioned graphs (Section~\ref{subsec:cons_and_access}). Partition access has an ORAM access complexity of $O(PT\log_{P}(\frac{b}{T}))$ or $O(PT\log_{P}(\frac{b^2}{T}))$ for \voram and \eoram, respectively (Section~\ref{sec:engine}). Partition processing is linear to the partition size (Section~\ref{subsec:basic_query}).
The discussion of complexities on \mat and \elist is left to Section~\ref{subsec:setup}, and Appendix~\ref{app:complexity} provides the detailed derivation. 

\tinypara{Partition size $l$ analysis. } We analyze the distributions of the partition size $l$ across five distributed graphs using the \emph{Monte Carlo} method. Our findings in Section~\ref{subsec:eval-performance} indicate that, across all distributions, the partition size $l$ lies in a small range and is notably smaller than the total edge count, \ie $l \ll |E|$, indiating that \sysname can effectively process queries by accessing only one partition. The worst-case partition size $l=|E|$ is \emph{unlikely} in practice because it requires: (1) all the edges in the graph only involve at most $2k$ vertices, $k$ for starting vertices and $k$ for the endings, and (2) the graph forms a bipartite graph starting from and ending in two chunks after the \emph{random} permutation in the initialization stage (Section~\ref{sec:engine}).

%The worst-case partition size $l=|E|$ happens when all the edges are in the same partition, which only happens for graphs where all the edges start from and end in $k$ vertices, which is rare in practice. Also, the starting and ending vertices are allocated into two chunks after the random permutation in the construction stage (Section~\ref{sec:engine}) when $k>1$, whose probability is negligible in practice. 
% \fanxy{or just delete...}

\section{Evaluation}\label{sec:eval}

We evaluate \sysname on a variety of graphs and queries to demonstrate its scalability and performance.

\subsection{Evaluation Setup}\label{subsec:setup}

\tinypara{Setup.} We implement the \sysname prototype based on ABY3~\cite{mohassel2018aby3}, a popular 3-party MPC platform.  We use three computation servers on the cloud, each equipped with 16 $\times$ 2.0GHz Intel CPU cores, 512GB memory and 10Gbps full duplex Ethernet with an average round-trip-time (RTT) of 0.12 ms.  Note that we use the entire memory just to support a larger-scale \mat baseline. 

\tinypara{Workload.} 
We use graphs of two types over all five queries in Section~\ref{sec:local_query}.  First, we use three open-source real-world graphs, including Slashdot~\cite{slashdot}, DBLP~\cite{dblp}, and Twitter~\cite{twitter}, with the number of edges ranging from less than 1 million to more than 1 \emph{billion}.  Figure~\ref{fig:perf_realworld} in Section~\ref{subsec:eval-application} lists the key parameters of these graphs.
To show the detailed performance characteristics of \sysname, we also use thirty synthetic graphs of five different edge distributions from \code{igraph}~\cite{igraph}.  For each distribution, we range the number of vertices $|V|$ from 1K to 32K. Table~\ref{tab:igraphs} in Section~\ref{subsec:eval-performance} summarizes the key parameters of these graphs.

\tinypara{Strawman solutions for comparison.} 
Since there are no systems known to us that can provide the same functionality as \sysname, we implement the strawman solutions in Section~\ref{subsec:motivation}, upon which we also implement the same five queries. 
We fine-tune the performance to add vectorization and parallelism, wherever applicable, making sure that the performance difference from implementation quality is minimized. 
In fact, much of the low-level code is shared with \sysname, including the ORAM and all the secure protocols.

For \mat, we first implement the three basic queries as follows: 
1) \code{EdgeExist} query on edge $(\mathsecret{v_i}, \mathsecret{v_j})$: we directly extract the element $(i, j)$ using the adj-\eoram (Section~\ref{subsec:motivation}) and compare it to $0$; 
2) \code{NeighborsCount}: we access adj-\voram and sum up the number of edges; 
and 3) \code{NeighborsGet}: we access adj-\voram, and compare the elements with $0$ through \prot{GT} operator to hide the exact number of edges, and then return the result to the client.
The \code{CycleIdentify} is then implemented by composing six \code{EdgeExist}, the same as \sysname. We ignore the \code{NeighborsFilter} query because \mat can not support multi-graphs, as we analyzed in Section~\ref{subsec:motivation}.
For \elist, we implement all queries by scanning the whole edge list, reusing \sysname's implementation on each partition.
% with some manual tuning on choosing the right batching and parallelism parameters.  

\tinypara{Execution time measurements.}
All the reported execution times are the wall-clock time measured on the computation servers from the start to the end of the initialization or query processing. We report the average from 5 runs. 
For \sysname and \mat that require ORAM accesses, the query processing time is the averaged time of successive $T$ queries\footnote{For static queries, we can initialize fresh ORAMs in background processes and directly use a fresh ORAM when the stash is full.}, where $T$ is the stash size, and we use $T = \sqrt{\text{\#(items in ORAM)}}$, the default setting in ~\cite{zahur2016revisiting}.

\begin{figure*}[t]
    \centering
    \includegraphics[width=0.9\textwidth]{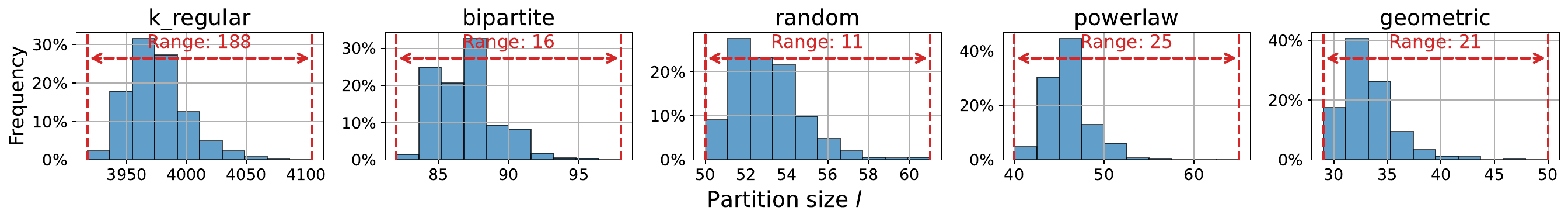}
    % \vspace{-.13in}
    \caption{Distribution of Partition Size $l$ on Varied Distributed Graphs ($|V| = 32K$).}
    \label{fig:l_distribution}
    % \vspace{-.1in}
\end{figure*}
  
\subsection{Performance on Real-world Graphs}
\label{subsec:eval-application}
\label{sec:eval-parallel}

We first provide an overview of \sysname performance using three real-world graphs with 700K to 1.4 billion edges.

%Our experiment shows that with parallel, we can minimize  While initializing the 

%Figure~\ref{fig:parallel_goram} shows the parallel initialization cost. Specifically, we split the $b \times b \times l$ 2d-partitioned 

%By leveraging $p=16$ processes, we can construct both \voram and \eoram for the billion-edge-scale graph within 2.9 minutes. This achieves a speedup of $9.4\times$ compared to the sequential construction. 
%We can not achieve the optimal $p\times$ speedup because 10Gbps bandwidth limits the cross-party communications.

%For the largest Twitter, \sysname can accelerate its performance through parallelization, as analyzed in Section~\ref{sec:eval-parallel}. 

\tinypara{Overall Query Performance.  }
The first row of Figure~\ref{fig:perf_realworld} shows the query execution time for five queries.
The configuration parameters $k$ and $\bar{l}$ are the default values stated in Section~\ref{sec:engine}. 
For each query, we use the maximum vector size. We use single-threading for smaller graphs and 16 threads for the larger Twitter graph.

For smaller Slashdot~\cite{slashdot} and DBLP~\cite{dblp}, \sysname completes all the queries within 135.7 ms.  To our knowledge, this is the only system that supports sub-second ego-centric queries on these graphs with strong privacy guarantees.  
In comparison, \mat gets out of memory even with the entire 512GB of memory per server, and \elist is $15.9\times$ and $4.2\times$ slower on average on these two graphs.

For the large Twitter, \sysname only takes 58.1 ms to 35.7 seconds to complete all queries, achieving a remarkable average speedup of $473.5\times$ over \elist.
Across all queries, we find that \sysname achieves more significant speedups for \code{EdgeExist} and \code{CycleIdentify}, with $856.3\times$ and $1445.6\times$ speedups, respectively.  This is because these two queries use the \eoram, with which we only need access \emph{one} of the $b^2 = 4096$ edge blocks with 1.1M edges, which is less than $0.8\permil$ of the total edges.
Instead, the other three use the \voram that accesses one \emph{row} of the $b=64$ edge blocks, which contain 71.2M edges, accounting $4.9\%$ for the total edges.

\tinypara{Fast query execution comes from processing less data.}
To verify that the execution time reduction indeed comes from reduced computation through partitioning, we also measure the total bytes transferred in each query on each graph. The second row in Figure~\ref{fig:perf_realworld} shows the results, which are consistent with execution time -  we observe an average communication reduction of $78.4\%$ compared to \elist. Also, the maximum reduction of $99.9\%$ is observed in \code{EdgeExist} and \code{CycleIdentify}, for the same reason above. 

\tinypara{Fast query execution also comes from parallel execution.  }
Query performance on large graphs also benefits from parallelization, as shown in Figure~\ref{fig:parallel}. On the large Twitter, we observe that using 16 threads, we can achieve an average speedup of $6.3\times$ across the five queries over a single thread.
We observe that except for \code{NeighborsGet}, we can achieve almost linear scalability using up to 16 threads. The slight derivation from linear scalability is because 8 threads already accelerate the computation to sub-second levels; adding more threads provides minimal speedup while increasing the aggregation overhead.
\code{NeighborsGet} does not parallel well because the \prot{SHUFFLE} procedure (line 8 in Algo.~\ref{algo:neighbors_get}) needs to process the entire partition to permute the result, which is inherently sequential.

\tinypara{Initialization performance.}
Unlike \elist, both \voram and \eoram require a non-trivial initialization step to construct the secure indices (Section~\ref{subsec:cons_and_access}).
While the two smaller graphs only take a few seconds, it takes dozens of minutes on the large Twitter using a naive sequential algorithm.
However, we can parallelize the initialization using multiple threads, as shown in Section~\ref{subsec:parall}. 
In this way, we can construct both \voram and \eoram for the billion-edge-scale graph within $2.9$ minutes using 16 threads.  
Figure~\ref{fig:parallel_goram} shows the experiments using different numbers of threads.
We observe an average speedup with 16 threads is $9.4\times$ over a single thread, and at this setting, we have saturated the 10Gbps network bandwidth.

\subsection{Micro-benchmarks}\label{subsec:eval-performance}

\begin{table}[t]
% \footnotesize
\caption{Synthetic Graphs}\label{tab:igraphs}
\vspace{-.1in}
\resizebox{0.4\textwidth}{!}{
\renewcommand{\arraystretch}{1}
\begin{tabular}{c|l|c}
\midrule[1.1pt]
Graph Types      & Generation Methods                        & Average Degree \\ \hline \hline
\code{k\_regular} & \code{K\_Regular}~\cite{igraphKR}        & 7.5                                \\ \hline
\code{bipartite} & \code{Random\_Bipartite}~\cite{igraphRB} & 134.4                              \\ \hline
\code{random}    & \code{Erdos\_Renyi}~\cite{igrphRAN}      & 268.8                              \\ \hline
\code{powerlaw}  & \code{Barabasi}~\cite{igraphBara}        & 523.7                              \\ \hline
\code{geometric} & \code{GRG}~\cite{igraphGEO}              & 1198.5                             \\ \hline
\end{tabular}
}
% \vspace{-.1in}
\end{table}

\begin{table}[t]
\small
\caption{Parameters of Synthetic Graphs with $|V|$=32K}\label{tab:graph_para}
\vspace{-.1in}
\resizebox{0.43\textwidth}{!}{
\renewcommand{\arraystretch}{1}
\begin{threeparttable}
\begin{tabular}{c|c|c|c|c|c|c}
\midrule[1.1pt]
Graph   Type  & $|E|$ & $k$ & $b$  & $l$  & $bl$  & ${b^2l} /   {|E|}$ \\ \hline\hline
\code{k\_regular} & 0.2M & 4096  & 8    & 3960 & 31.8K & 1.03               \\ \hline
\code{bipartite}  & 13.4M & 64 & 512  & 88   & 44.5K & 1.72               \\ \hline
\code{random}     & 26.8M & 32 & 1024 & 56   & 57.3K & 2.19               \\ \hline
\code{powerlaw}   & 52.3M & 16 & 2048 & 48   & 98.3K & 3.85               \\ \hline
\code{geometric}  & 105.0M & 8 & 4096 & 32   & 131.1K  & 5.11               \\ \hline
\end{tabular}
\begin{tablenotes}
    \item {\footnotesize { The parameters $k$, $b$ and $l$ refer to \#(vertices per chunk), \#(vertex chunks) and the edge block size with padded edges, see Section~\ref{sec:engine}. $l$ and $bl$ are the {\sf EORAM} and {\sf VORAM} partition sizes, respectively. }}
\end{tablenotes}
\end{threeparttable}
}
\vspace{-.1in}
\end{table}

\begin{figure*}[t]
	\centering
	\includegraphics[width=0.9\linewidth]{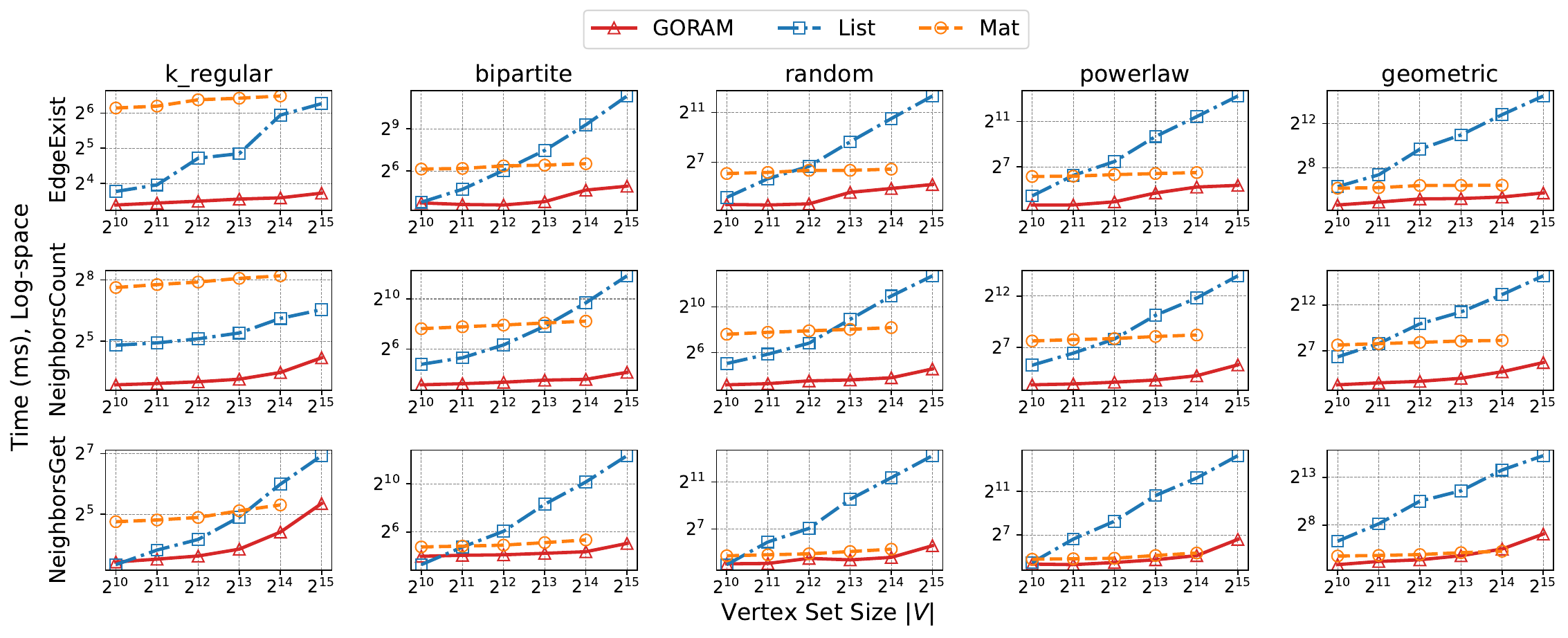}
	\vspace{-.1in}
	\caption{Online Performance Overview}
	\label{fig:online_performance}
\end{figure*}

\begin{figure*}[t]
	\centering
	\includegraphics[width=0.9\linewidth]{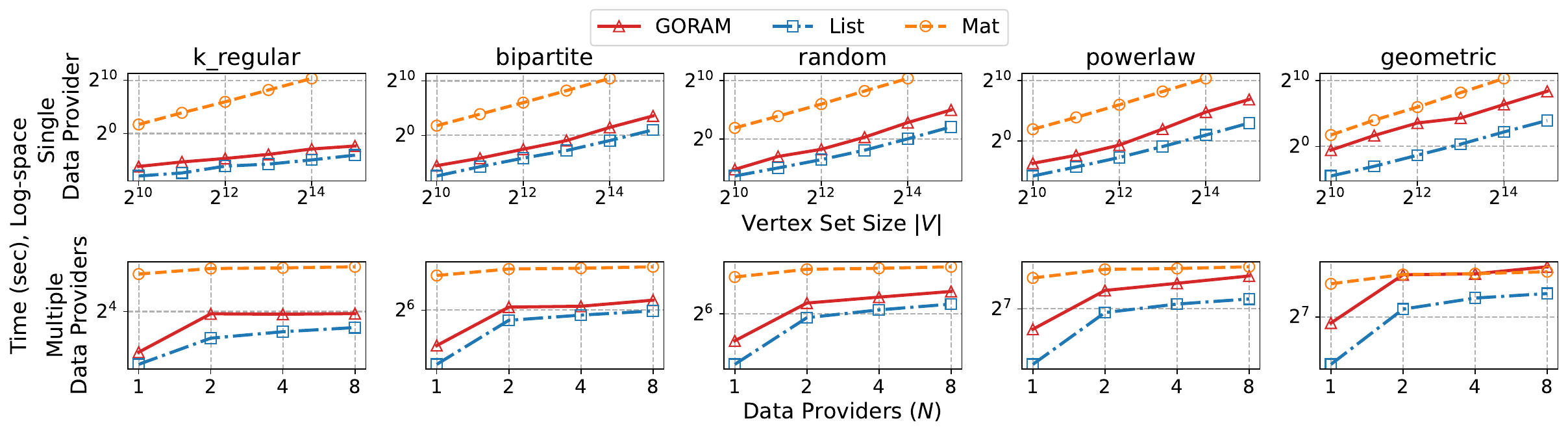}
	\vspace{-.1in}
	\caption{Initialization Cost}
	\label{fig:offline_single}
\end{figure*}

To better understand \sysname performance, we run queries on 30 synthetic graphs of various distributions and sizes, measure the execution time, and compare to the two strawman solutions.  Limited by space, we only present three basic queries \code{NeighborsCount}, \code{NeighborsGet} and \code{EdgeExist}, because the other two are based on these queries and thus \code{CycleIdentify} offers similar performance as \code{EdgeExist}, and \code{NeighborsFilter} is similar to \code{NeighborsCount}.  
We focus on single-thread-only for all micro-benchmarks to best illustrate the performance benefit from the \sysname data structure. 

\tinypara{Adaption to various distributed graphs.}  Each distribution of the graph presents a different density, i.e. vertex degree $D = \frac{|E|}{|V|}$, with the average degree shown in Table~\ref{tab:igraphs}. Among them, the \code{k\_regular} is the sparsest and the \code{geometric} is the densest.  
We present the default configuration parameters $k$ of \sysname (see Section~\ref{sec:engine}) and the corresponding vertex chunk numbers $b$, the partition sizes for \eoram ($l$) and \voram ($bl$) in Table~\ref{tab:graph_para}. The ratio $\frac{b^2l}{|E|}$ is also shown, representing the amplification factor of the 2d-partitioned structure with padded edges for security compared to the original edge list.
We observe that for sparse \code{k\_regular}, \sysname uses $k$ as large as 4096, while for \code{geometric}, $k$ gets as small as 8. The choice of $k$ makes intuitive sense. 
Recall that a key property when we partition is that all outgoing edges of a vertex are within a single row of the 2d-partition. Thus, when the graph is sparse, we have the opportunity to partition it into fewer chunks, \ie smaller $b=\frac{|V|}{k}$, enjoying the benefits from quicker partition access time.
However, on a dense graph, we partition it into more chunks to make each partition smaller, thereby reducing the partition scanning time per query.

\tinypara{Partition size $l$ distributions} across diverse distributed graphs are shown in Figure~\ref{fig:l_distribution}, derived through the \emph{Monte Carlo} method. 
For each distribution, we generate $10K$ graphs, each with $32K$ vertices. For each graph, we randomly permute the vertices using different random seeds and construct the 2d-partitioned structure to obtain the partition size $l$ as Section~\ref{sec:engine} illustrates without padding.
As Figure~\ref{fig:l_distribution} shows, the range of $l$ is relatively narrow for all distributions. E.g., $l$ varies within a range of $188$ in \code{k\_regular} and only varies within a range of $11$ in \code{random} graph. 
Notably, the partition size $l$ is significantly smaller than $|E|$ for all five graph types (see Table~\ref{tab:graph_para}), suggesting that \sysname often significantly outperforms \elist.

% Form Figure~\ref{fig:l_distribution}, we can see that the range of $l$ is relatively narrow for all distributions. 
% For instance, in \code{k\_regular} graph, $l$ varies within a range of $188$, and in \code{random} graph, $l$ varies in a range of $11$. 
% Notably, for all five graph types, the partition size $l \ll |E|$. Therefore, for most cases, \sysname outperforms \elist with significant advantages.

% To push it to the extreme, when the graph is a complete graph, \sysname degrades to \mat ($k=1$, or $|V|$ chunks)
% Thus, when the graph is sparse, we have the opportunity to partition it into fewer chunks, enjoying the gain from quicker partition access time and correspondingly smaller chunk processing per query.  
% However, on a dense graph, we partition it into more chunks to make each partition smaller .  
% To push it to the extreme, when the graph is a complete graph, \sysname degrades to \mat ($k=1$, or $|V|$ chunks).  

\tinypara{Query execution time.}
Each row of Figure~\ref{fig:online_performance} presents the execution time of a query across different graph densities from the sparsest to the densest, using both \sysname and the two strawman solutions.  
Overall, we observe that \sysname delivers highly efficient query responses across all 90 test cases (6 sizes $\times$ 5 graph distributions $\times$ 3 queries), offering an average query completion time of $22.0$ ms.  We provide a detailed performance analysis below.

\tinypara{Performance vs. graph density and size.}
\mat is the least sensitive to graph density (e.g., all around $2^6$ ms across the first row), because its query processing time only depends on $|V|$ (see Table~\ref{tab:complexity}). 
\elist, on the contrary, is very sensitive to density, given its $O(|E|)$ processing time per query. 
\sysname, on the other hand, works well across different densities, exhibiting sub-linear execution time on both $|V|$ and the graph density.  The trend in execution time precisely matches the theoretical complexities in Table~\ref{tab:complexity}.  

For sparse graphs (first two columns in Figure~\ref{fig:online_performance}), \mat performs the worst on almost all $|V|$ settings, as expected, because it spends too much resource processing empty cells.  
\elist performs as well as, or even better than \sysname on very small graphs (\ie 1024 vertices) for \code{NeighborsGet}. 
This is because \code{NeighborsGet} requires multiple secure comparison and multiplication operations, and the communication round latency becomes the bottleneck for small graphs. \elist, in this case, saves the communication rounds required for partition access, providing advantages compared to \sysname.
However, the performance gets worse fast as $|V|$ increases, because the $O(|E|)$ complexity of \elist quickly dominates the performance.

For dense graphs (last two columns in Figure~\ref{fig:online_performance}), \mat gets closer performance with \sysname as far as it supports the scale, but \elist performs poorly except for the smallest cases.  In fact, in the largest version of the densest \code{geometric} graph, \elist can be as much as $703.6\times$ slower because of its $O(|E|)$ complexity.

\tinypara{Performance with queries.}
\mat is always several times slower than \sysname on \code{NeighborsGet}, because it takes $O(\log(|V|))$ time to access the adj-\voram while \sysname only takes $O(\log(b))$, $b = \frac{|V|}{k}$. 

For \code{EdgeExist}, although \sysname still requires less ORAM access time, the advantage is less significant because it introduces $O(\log(l))$ communication rounds of \prot{OR} to aggregate the partition (see Section~\ref{subsec:basic_query}). However, \mat only needs a single round of comparison to process a single cell. Therefore, \mat becomes closer to \sysname for denser graphs with larger $l$.

The slowest query of \sysname is the \code{NeighborsGet} on the largest \code{geometric} graph, which takes $132.8$ ms. The performance of \sysname and \mat are very close except for the sparsest \code{k\_regular}. This is because \sysname requires multiple expensive secure operations on the vertex partition with $bl$ edges while \mat only needs a single comparison on $|V|$ matrix cells, which becomes faster than \sysname for denser graphs, offsets the higher overhead of ORAM access, \ie $O(\log(|V|))$ for \mat while $O(\log(b))$ for \sysname.

% Except for the sparsest \code{k\_regular}, the performance is closer to that of \mat.  This is because \code{NeighborsGet} requires more secure operations than the other two queries and the partition process becomes the bottleneck. Given that the \sysname partition size $bl$ becomes larger than $|V|$ with denser graphs, the performance of \sysname gets closer to \mat.

\tinypara{Initialization performance.} 
The first row in Figure~\ref{fig:offline_single} shows the initialization cost when there is only one data provider. The cost is the wall-clock time from data loading to secure indices construction (affecting only \sysname and \mat, as \elist does not require establishing indices). The initialization cost is linear to the graph sizes, \ie $|V|^2$ or $|E|$. 
Given the relationship $|E| < b^2l < |V|^2$ among \elist, \sysname, and \mat, the initialization costs follow the order: \mat $>$ \sysname $>$ \elist, as shown in Figure~\ref{fig:offline_single}.
\mat has the highest and constant cost across sparse and dense graphs because of its $O(|V|^2)$ complexity. Notably, \mat runs out of memory during the construction of adj-\eoram on all graphs with 32K vertices, even with 512GB memory.

\tinypara{Initialization performance with multiple data providers.}  The second row in Figure~\ref{fig:offline_single} presents the cost when there are multiple data providers (1 to 8). We simulate the distributed graphs by randomly assigning each edge of the synthetic graph with 16K vertices to each data provider, which is the largest scale \mat supports. 
For $N>1$ data providers, both \sysname and \elist need to perform the merge sort on $N^{+}$ or $N$ ordered private graphs (see Section~\ref{subsec:cons_and_access}), which leads to more overhead. However, the impact of an increase in $N$ is not significant as Figure~\ref{fig:offline_single} shows. This is because the introduced workload is logarithmic, \ie $\log(N)$ or $\log(N^{+}), N^{+} \approx \frac{l}{\bar{l}}$.

\begin{figure}[t]
    \centering
    \includegraphics[width=0.47\textwidth]{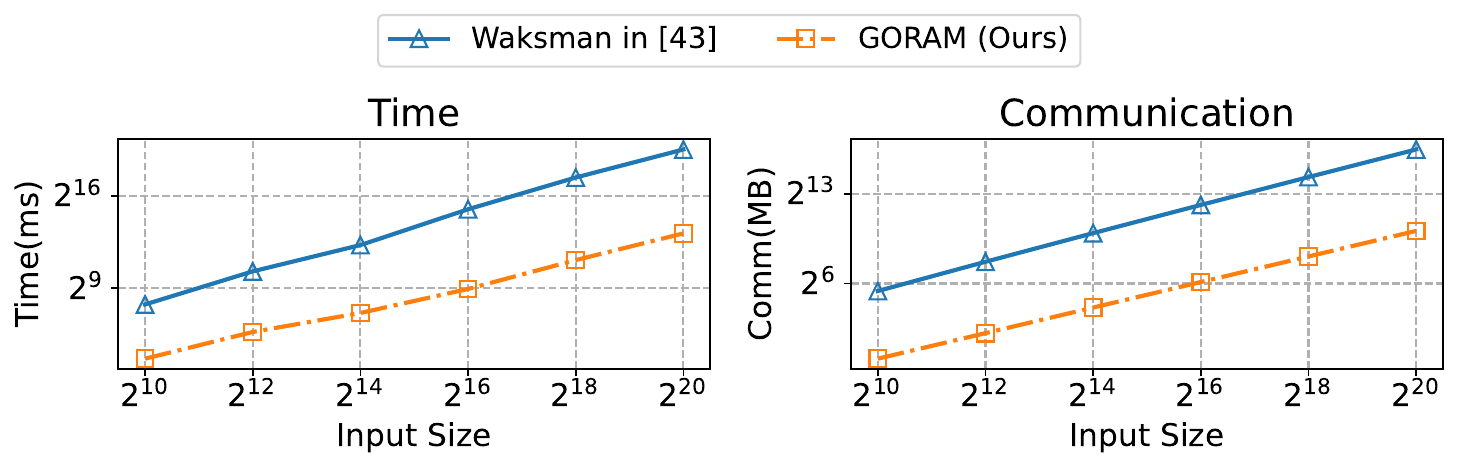}
    \vspace{-.1in}
    \caption{\prot{ShuffleMem} Construction {\footnotesize (* the y-axes are in log-scale. )}}
    \label{fig:eval-permutation}
    \vspace{-.13in}
\end{figure}

\tinypara{ShuffleMem construction comparison.} Figure~\ref{fig:eval-permutation} compares the cost of \prot{ShuffleMem} construction of an array of $n$ secret-shared integers using \emph{Waksman} permutation network, as adopted in \cite{zahur2016revisiting}, and our optimized constant-round \prot{ShuffMem} protocol introduced in Section~\ref{subsec:shuffmem}. The \prot{ShuffleMem} construction is the main bottleneck of building ORAM. 
We can see that \sysname significantly accelerates both computation and communication, achieving $17.4\times$ to $83.5\times$ speedups and $97.5\%$ to $98.8\%$ communication savings as input sizes increase. 
This is because our method reduces the original $O(n\log(n))$ computation and communication to $O(n)$, thereby showing better performance as input sizes increase. 
Furthermore, unlike Waksman network, which necessitates an expensive switch operation, amounting to approximately $\approx 6n$ communications per layer in the $2\log(n)$ depth network, \sysname only requires shares transmission and \prot{XOR} operations that do not need communications.

% GORAM only requires shares transmission and \prot{XOR} operations, unlike the Waksman network, which necessitates an expensive switch operation, amounting to approximately $\approx 6n$ communications per layer in the $2\log(n)$ depth network.

\section{Related Work}\label{sec:rela_work}

We discuss the prior arts related to \sysname.

\tinypara{Secure federated databases} focus on conducting \emph{public} SQL queries over federated databases while protecting the individual tuples. Examples include~\cite{bater2017smcql, bater2018shrinkwrap, volgushev2019conclave, bater2020saqe, han2022scape, liagouris2023secrecy}.
For each query, these databases analyze the statements and run secure protocols on the required data to obtain the result. 
Based on the above progress, Aljuaid \etal~\cite{al2023secure, aljuaidefficient} propose to process federated graph queries by directly translating the graph queries into SQL through~\cite{jindal2014vertexica}.
Compared with these \emph{public} query systems, we focus on ego-centric queries with private query keys.

\tinypara{Secure graph processing with theoretical guarantees.} 
Beyond the \mat and \elist introduced in Section~\ref{subsec:motivation}, there are proposals leveraging \emph{Structured Encryption (SE)}~\cite{chase2010structured} to query graphs securely.
They focus on encrypting the graph in a way that can be privately queried~\cite{9865984, lai2019graphse2, meng2015grecs, falzon2024pathges}.
However, they require a shared key between the data provider and the client, and consequently, cannot be directly extended to allow multiple data providers and third-party clients.
Other studies~\cite{liu2024federated,roth2021mycelium} focus on private analytics over a set of devices organized as a graph, a topic orthogonal to our work.

\tinypara{Graph processing under other security settings.}
Numerous proposals focus on conducting graph queries guaranteeing \emph{differential privacy (DP)} on two neighboring graphs, \eg graphs differ on one vertex or edge, like ~\cite{mazloom2018secure,mazloom2020secure,imola2021locally, karwa2011private, liu2024federated,roohi2019differentially,10113303,299537}.
They focus on protecting inputs from the results. In addition to the unavoidably inaccurate results, they pay less attention to the information leakage during the computation. E.g., FEAT~\cite{liu2024federated} leaks noisy vertex degrees, which may be close approximations to control errors. \sysname, however, focuses on protecting \emph{all} the information during the computation except the result, which is orthogonal to DP's goal. 
There are proposals leveraging TEEs~\cite{chamanigraphos, xu2023framework,chang2016privacy, chang2022towards}, which are vulnerable to side-channel attacks~\cite{lou2021survey, munoz2023survey}.

\tinypara{DORAM implementations. }
FLORAM~\cite{falk2023gigadoram} and DuORAM~\cite{vadapalli2023duoram} focus on high-latency and low-bandwidth settings. They trade a linear computation complexity for reduced communications, therefore becoming impractical for large-scale data.
3PC-DORAM~\cite{bunn2020efficient}, GigaORAM~\cite{falk2023gigadoram}, and Square-root ORAM~\cite{zahur2016revisiting} struggle for sub-linear complexity. 
GigaORAM and 3PC-DORAM depend on the \emph{Shared-In Shared-Out Pseudo Random Functions (SISO-PRF)} to improve complexity. However, the SISO-PRF becomes practical only with ``MPC friendly'' block ciphers, \ie LowMC~\cite{albrecht2015ciphers}, which was unfortunately cryptanalyzed~\cite{liu2021cryptanalysis}. 
\sysname builds its indices on Square-root ORAM, ensuring sublinear complexity and robust security guarantee.

\section{Conclusion and Future Work}

We propose \sysname, the first step towards achieving efficient private ego-centric queries on federated graphs. 
\sysname introduces a methodology for reducing the to-be-processed data sizes in secure computations, relying on query-specific data partitioning and secure indices. We hope this method can be generalized to other applications beyond ego-centric queries.
Extensive evaluations validate that \sysname achieves practical performance on real-world graphs, even with 1.4 billion edges.
For future work, we aim to expand \sysname's capabilities for more advanced applications, including complex graph queries like path filtering and pattern matching, while also optimizing its performance and scalability.
% For future work, we plan to extend \sysname to support other advanced applications, including complex graph queries like path filtering and sub-graph pattern matching. We will also optimize its performance and scalability. 
% We will open source \sysname with its integration on ABY3 upon publication of this paper.

\clearpage
\bibliographystyle{ACM-Reference-Format}
\bibliography{main}

% \clearpage
\appendix

\section{Security Analysis}\label{app:security}

\subsection{Security Proof of ShuffleMem}\label{app:shuffle-sec}

In this section, we provide a formal proof of Theorem~\ref{thm:sec-shufflemem}, following strictly to the \emph{real-world / ideal-world paradigm}~\cite{canetti2000security}. Let $\mathcal{A}$ denote the real-world adversary, and $\mathcal{S}$ denote the ideal-world adversary. We prove that for \emph{any} semi-honst $\mathcal{A}$ controlling at most one party, there exists a $\mathcal{S}$ running in polynomial of the running time of $\mathcal{A}$, such that for all inputs and all sets of corrupted parties, the view of $\mathcal{S}$ is indistinguishable from the view of $\mathcal{A}$ (Equation~\ref{eq:sec-view}). 

We prove the security of Protocol~\ref{prot:shuffleMem} in the $\mathcal{F}_{\text{setup}}$-hybrid world, where there exists an ideal functionality $\mathcal{F}_{\text{setup}}$ that generates common \emph{pseudorandom keys (PRF)} among the computation parties, \ie each pair of computation servers, $S_i$ and $S_j$, has a common key $s_{ij}$ to generate correlated randomness, and the key is exclusively known to the pair of computation servers. 
The security of protocols implemented $\mathcal{F}_{\text{setup}}$ is well-established, and \cite{koti2024mathsf} provides the formal proof; therefore, we omit it here for brevity. 

In the following, we define the ideal functionality of \prot{ShuffleMem}, \ie $\mathcal{F}_{\prot{ShuffleMem}}$, and then provide the construction of the simulator $\mathcal{S}$ to prove the security of Protocol~\ref{prot:shuffleMem}.

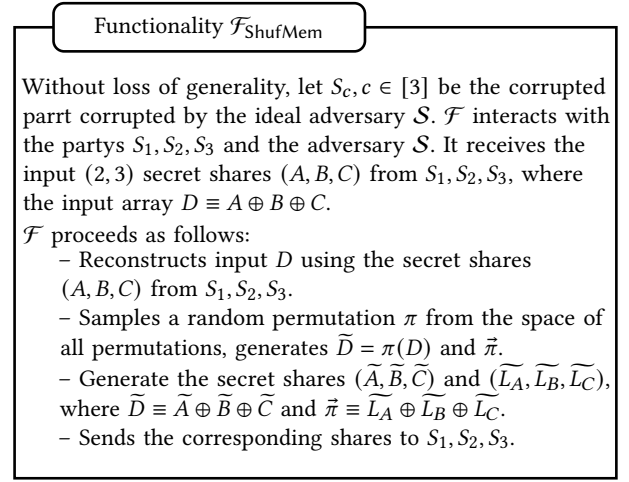
\begin{figure}[t]
    \begin{tikzpicture}
        \def\boxwidth{8cm}
        \def\boxheight{6cm}
    
    % 绘制主外框
    \draw [line width=1pt] (0,0) rectangle (\boxwidth,-\boxheight);

    \node[anchor=west, draw, line width=1pt, inner sep=2pt, fill=white, rounded corners] at (0.5,0) {
        \begin{minipage}[c][0.5cm][c]{\boxwidth-4cm}
            \centering
            Functionality $\mathcal{F}_{\sf{ShufMem}}$
        \end{minipage}
    };
    
    % 这里可以根据需要绘制分割线，比如 Preprocessing、Online 的分隔线
    % 下面的例子为在框中绘制水平线分隔段落
    %   \draw [line width=0.5pt] (0,-1) -- (\boxwidth,-1);
    
    % 标题文字
    \node[anchor=west, text width=\boxwidth-0.2cm] at (0,-1.5) {
        
    Without loss of generality, let $S_c, c \in [3]$ be the corrupted parrt corrupted by the ideal adversary $\mathcal{S}$. $\mathcal{F}$ interacts with the partys $S_1, S_2, S_3$ and the adversary $\mathcal{S}$. It receives the input $(2,3)$ secret shares $(A, B, C)$ from $S_1, S_2, S_3$, where the input array $D \equiv A \xor B \xor C$. 
    
    };
    
    \node[anchor=west, text width=\boxwidth-0.2cm] at (0, -2.7) {

        $\mathcal{F}$ proceeds as follows:

    };

    \node[anchor=west,text width=\boxwidth-0.7cm] at (0.5,-4.3) 
    {
    – Reconstructs input $D$ using the secret shares $(A, B, C)$ from $S_1, S_2, S_3$.

    – Samples a random permutation $\pi$ from the space of all permutations, generates $\widetilde{D} = \pi (D)$ and $\vec{\pi}$.

    – Generate the secret shares $(\widetilde{A}, \widetilde{B}, \widetilde{C})$ and $(\widetilde{L_A}, \widetilde{L_B}, \widetilde{L_C})$, where $\widetilde{D} \equiv \widetilde{A} \xor \widetilde{B} \xor \widetilde{C}$ and $\vec{\pi} \equiv \widetilde{L_A} \xor \widetilde{L_B} \xor \widetilde{L_C}$.

    – Sends the corresponding shares to $S_1, S_2, S_3$.
    };

    \end{tikzpicture}
    \caption{Ideal Functionality $\mathcal{F}_{\sf{ShufMem}}$. }
    \label{fig:functionality}
\end{figure}

The definition of $\mathcal{F}_{\prot{ShuffleMem}}$ is shown in Figure~\ref{fig:functionality}, which is defined as a trusted third party that randomly shuffles the secret shared array $D \equiv A \xor B \xor C$ using a random permutation $\pi$ and returns the secret shares of the shuffled array $\widetilde{D} = \pi(D)$ and the permutation representation $\vec{\pi}$ to the corresponding computation servers. 
Then, we prove that the real-world protocol $\Pi_{\prot{ShuffleMem}}$ (Protocol~\ref{prot:shuffleMem}) securely realizes $\mathcal{F}_{\prot{ShuffleMem}}$ in the presence of \emph{any} semi-honest adversary $\mathcal{A}$. 
Without loss of generality, the simulator is constructed assuming $\mathcal{A}$ corrupts the party $S_1$. The simulation for $S_2$ and $S_3$ is similar. 

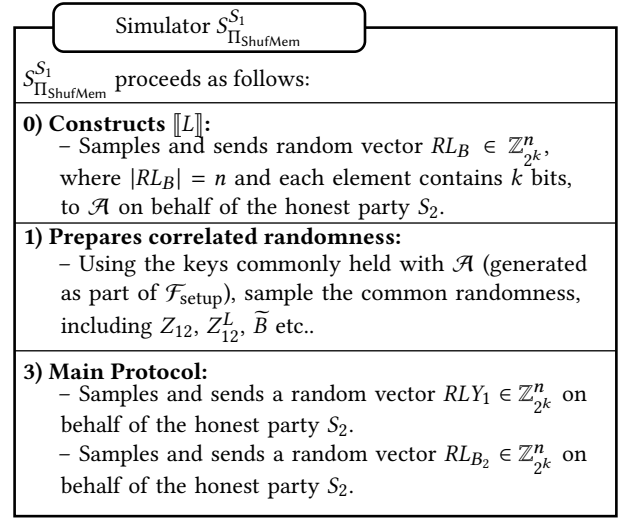
\begin{figure}[t]
\begin{tikzpicture}
  \def\boxwidth{8cm}
  \def\boxheight{6.5cm}

  % 绘制主外框
  \draw [line width=1pt] (0,0) rectangle (\boxwidth,-\boxheight);

    \node[anchor=west, draw, line width=1pt, inner sep=2pt, fill=white, rounded corners] at (0.5,0) {
        \begin{minipage}[c][0.5cm][c]{\boxwidth-4cm}
            \centering
            Simulator $S_{\Pi_{\sf{ShufMem}}}^{S_1}$
        \end{minipage}
    };

  % 这里可以根据需要绘制分割线，比如 Preprocessing、Online 的分隔线
  % 下面的例子为在框中绘制水平线分隔段落
  \draw [line width=0.5pt] (0,-1) -- (\boxwidth,-1);

  % 标题文字
  \node[anchor=west] at (0,-0.7) {$S_{\Pi_{\sf{ShufMem}}}^{S_1}$ proceeds as follows:};

  \node[anchor=west,font=\bfseries] at (0,-1.3) {0) Constructs $\mathsecret{L}$:};
  \node[anchor=west,text width=\boxwidth-1cm] at (0.5,-2.0) 
  {
    – Samples and sends random vector $RL_B \in \mathbb{Z}^{n}_{2^{k}}$, where $|RL_B| = n$ and each element contains $k$ bits, to $\mathcal{A}$ on behalf of the honest party $S_2$.
  };

  % Preprocessing段落标题
  \draw [line width=0.5pt] (0,-2.6) -- (\boxwidth,-2.6);
  \node[anchor=west,font=\bfseries] at (0,-2.8) {1) Prepares correlated randomness:};

  % Preprocessing 内容项 
  \node[anchor=west,text width=\boxwidth-1cm] at (0.5,-3.6) 
  {
    – Using the keys commonly held with $\mathcal{A}$ (generated as part of $\mathcal{F}_{\text{setup}}$), sample the common randomness, including $Z_{12}$, $Z^{L}_{12}$, $\widetilde{B}$ \etc.
  };

  % 再次绘制水平线，分隔到 Online 阶段
  \draw [line width=0.5pt] (0,-4.3) -- (\boxwidth,-4.3);

  % Online段落标题
  \node[anchor=west,font=\bfseries] at (0,-4.6) {3) Main Protocol:};

  % Online 内容项
  \node[anchor=west,text width=\boxwidth-1cm] at (0.5,-5.5)
  {
    – Samples and sends a random vector $RLY_1 \in \mathbb{Z}^{n}_{2^{k}}$ on behalf of the honest party $S_2$.

    – Samples and sends a random vector $RL_{B_2} \in \mathbb{Z}^{n}_{2^{k}}$ on behalf of the honest party $S_2$.
  };
  \end{tikzpicture}
  \caption{Simulator $S_{\Pi_{\sf{ShufMem}}}^{S_1}$ Construction. }
  \label{fig:simulator}
\end{figure}

% \begin{figure*}[t]
%     \centering
%     \includegraphics[width=1.0\textwidth]{./graph/partition_size_distribution.pdf}
%     \caption{Distribution of Partition Size $l$ on Varied Distributed Graphs ($|V| = 32K$).}
%     \label{fig:l_distribution}
% \end{figure*}

\begin{proof}
    Let $\mathcal{A}$ denote the real-world adversary corrupting $S_1$, and $\mathcal{S}_{\Pi_{\sf{ShufMem}}}^{S_1}$ denote the corresponding ideal-world adversary. The construction of $\mathcal{S}_{\Pi_{\sf{ShufMem}}}^{S_1}$ is shown in Figure~\ref{fig:simulator}.

    For the 0) step in Protocol~\ref{prot:shuffleMem}, the simulator $S_{\Pi_{\sf{ShufMem}}}^{S_1}$ samples and sends a random vector $RL_B \in \mathbb{Z}^{n}_{2^{k}}$ to $\mathcal{A}$ (since $\mathcal{A}$ corrupts $S_1$) on behalf of the honest party $S_2$. Because the real-world message $L_B = Z_{2}$ is uniformly random in the view of $\mathcal{A}$, $\mathcal{A}$ cannot distinguish the real-world messages $L_B$ with the randomly sampled message $RL_B$ from the simulator.

    For the 1) step, $\mathcal{S}$ emulates $\mathcal{F}_{\text{setup}}$ at first, during which $\mathcal{S}$ generates the PRFs between the corrupted $S_1$ and $S_2$, $S_3$, \ie $s_{12}, s_{13}$.  Then, $S$ generates the correlated randomness held by $S_1$, including $Z_{12}, Z^{L}_{12}, \widetilde{B}$ \etc.

    For the 3) step, $\mathcal{S}$ randomly samples $RLY_1$ and $RL_{B_2}$ on behalf of $S_2$ and sends to $\mathcal{A}$. In the real-world scenario, the values $LY_1$ and $L_{B_2}$ received by $\mathcal{A}$ are masked using $Z^{L}_{23}$ and $\widetilde{L}_C$, which are uniformly random from $\mathcal{A}$'s viewpoint. Consequently, the messages $LY_1$ and $L_{B_2}$ appear uniformly random to $\mathcal{A}$ and are indistinguishable from the messages provided by the simulator.
    
    % Because $LY_1$ and $L_{B_2}$ that $\mathcal{A}$ receives in the real-world are masked using $Z^{L}_{23}$ and $\widetilde{L}_C$ that are uniformly random in the view of $\mathcal{A}$, the messages $LY_1$ and $L_{B_2}$ are uniformly random to $\mathcal{A}$ and cannot be distinguished with the messages send by $\mathcal{S}$.

    In summary, the view of $\mathcal{A}$ in the real-world (\ie interacting with honest servers $S_2$ and $S_3$) is indistinguishable from the view of the ideal-world (\ie interacting with a simulator $S_{\Pi_{\sf{ShufMem}}}^{S_1}$). Therefore, we have Theorem~\ref{thm:sec-shufflemem} holds.
     
\end{proof}

\section{Complexity Analysis}\label{app:complexity}

In this section, we analyze the complexities summarized in Table~\ref{tab:complexity}.

\para{Initialization. } The initialization of \mat and \elist are dominated by the adj-\voram, adj-\eoram constructions, and odd\_even\_merge\_sort networks, respectively. \sysname, however, includes both stages while each stage needs lower-complexity because of the partitioned data structure (see Section~\ref{sec:engine}).

For \mat, the initialization includes a linear-complexity \prot{ShuffleMem} and a logmatric-complexity position map construction. Since \mat requires a $|V|^2$ adjacency matrix each data provider, the computation complexity is $O(N|V|^2)$. The communication rounds include the $3$ rounds for the \prot{ShuffleMem} Protocol~\ref{prot:shuffleMem} (Section~\ref{subsec:shuffmem}) and $2\log_P(\frac{|V|}{T})$ or $2\log_P(\frac{|V|^2}{T})$ rounds for the position map construction of adj-\voram and adj-\eoram, respectively (Section~\ref{subsec:goram-pre}).

For \elist, the initialization complexity is the complexity of the {odd\_even\_merge\_sort} network for $N$ data providers. Because the complexity of the merge sort network is $O(n\log(n))$ and the communication rounds is the $O(\log(n))$, the complexity of sorting of $N$ lists with $|E|$ elements is $O(|E|\log(|E|)\log(N))$, and the communication rounds is $O(\log(|E|)\log(N))$.

For \sysname, the initialization contains both ORAM initialization over the partitions and odd\_even\_merge\_sort for each partition. Since each of the $b^2$ partitions can be processed in parallel, the merging round complexity is $\log(l)\log(N^{+})$, where $l$ is the partition size and $N^{+}$ is the number of all the submitted partitioned graphs. Therefore, the complexity to initialize \sysname is $O(b^2l\log(l)\log(N^{+}))$, where the $b^2$ sorting network is the main bottleneck. The communication rounds include $3$ for the \prot{ShuffleMem} Protocol~\ref{prot:shuffleMem}, $O(\log(l)\log(N^{+}))$ for the sorting network, and $2\log_P(\frac{b}{T})$ or  $2\log_P(\frac{b^2}{T})$ for the position map construction of \voram and \eoram, respectively.

\para{Partition Access. } For \mat and \sysname that utilize ORAM, the partition access complexity is the same as the ORAM access complexity. By substituting the $n$ term in the ORAM access complexity, \ie $O(PT\log_P(\frac{n}{T}))$, we have that the partition access complexity for \mat is $O(PT\log_P(\frac{|V|^2}{T}))$ and $O(PT\log_P(\frac{|V|}{T}))$ for adj-\eoram and adj-\voram, respectively. Similarly, the partition access complexity for \sysname is $O(PT\log_P(\frac{b^2}{T}))$ and $O(PT\log_P(\frac{b}{T}))$ because there are $b^2$ and $b$ partitions in the \eoram and \voram, respectively.

The \elist, however, does not contain ORAM, and the partition access complexity is $O(1)$, \ie directly access the whole list.

\para{Partition Processing for Basic Queries. } The complexities of the partition processing are basically the complexity of scanning the whole partition. 

For the edge-centric query, \code{EdgeExist}, the partition size of \mat, \elist and \sysname is $1$ (\ie a single element in the adjacency matrix), $|E|$ and $l$ (\ie a single partition); therefore, the complexity of this query is $O(1)$, $O(|E|)$ and $O(l)$, respectively. Because the aggregation operation \prot{OR} requires communication, therefore the round complexity is $O(1)$, $O(\log(N))$ and $O(\log(l))$ for \mat, \elist and \sysname, respectively.

For the vertex-centric queries, i.e., \code{NeighborsCount} and \code{NeighborsGet}, the partition size of \mat, \elist and \sysname is $|V|$ (\ie a single row in the adjacency matrix), $|E|$ and $bl$; therefore, the complexity of these queries is $O(|V|)$, $O(|E|)$ and $O(bl)$, respectively. Because the aggregation operation in \code{NeighborsCount} is communication-free \prot{ADD} and \code{NeighborsGet} does not need aggregation, the round complexity for all three data structures across the two queries is $O(1)$.

% \section{Partition Size Distribution}\label{app:l_distribution}

% Figure~\ref{fig:l_distribution} shows the distribution of the partition size $l$ across diverse distributed graphs, obtained through the Monte Carlo sampling method. 
% For each distribution, we generate a set of $10,000$ graphs with $32K$ vertices. For each graph, we randomly permute the vertices using different random seeds, construct the 2d-partitioned structure to obtain the partition size, $l$, as Section~\ref{sec:engine} illustrates. 
% Form Figure~\ref{fig:l_distribution}, we can see that the range of $l$ is relatively narrow for all graph types. For instance, in \code{k\_regular} graph, $l$ varies within a range of $188$, and in \code{random} graph, $l$ varies in a range of $11$. 
% Notably, for all five graph types, the partition size $l \ll |E|$. Therefore, for most cases, \sysname outperforms \elist with significant advantages. 

\begin{algorithm}[t]
	\small
	\SetKwInOut{Input}{\textbf{Inputs}}
	\SetKwInOut{Output}{\textbf{Output}}
	\caption{\small \code{GetPosBase}}\label{algo:prefix-access}
	\Input{\code{ORAM} in the last level containing $T$ blocks, $\mathsecret{i}$ denote the secret index in this level, $\mathsecret{\prot{fake}}$.}
	\Output{Physical index $p$. }
        \BlankLine
        \comm{\cfont{For $\mathsecret{s_2}$ without \secret{\prot{fake}} in ~\cite{zahur2016revisiting}}}
        $\mathsecret{\prot{notUsed}} \leftarrow \code{ORAM}.\mathsecret{\prot{Used}}$\;
        $\mathsecret{\prot{fZero}} \leftarrow \mathsecret{0, \prot{notUsed}_0, \prot{notUsed}_1, \dots, \prot{notUsed}_{T-2}}$ \;
        \For{$i \gets 0$ \textbf{to} $\lfloor \log_2(T) \rfloor$}{
            $s = 2^{i}$ denoting the stride\;
            $\mathsecret{\prot{fZero}}_{s:T} \leftarrow \prot{OR}(\mathsecret{\prot{fZero}}_{s:T}, \mathsecret{\prot{fZero}}_{0:T-s})$\;
        }
        $\mathsecret{\prot{fZero}}.{\prot{append}(\mathsecret{1})}$\;
        $\mathsecret{\prot{fZero}}_{0:T} \leftarrow \prot{XOR}(\mathsecret{\prot{fZero}}_{0:T}, \mathsecret{\prot{fZero}}_{1:T+1})$\;
        \comm{\cfont{Update \secret{\prot{Used}}}}
        \comm{\cfont{For $\mathsecret{s_1}$ without \secret{\prot{fake}} in ~\cite{zahur2016revisiting}}}
        $\mathsecret{s_1} \leftarrow \prot{EQ}(\text{expanded} \mathsecret{i}, [0, 1, \dots, T-1])$\;
        \comm{\cfont{Considering \secret{\prot{fake}}}}
        $\mathsecret{\prot{mask}} \leftarrow \mathsecret{s_1}\, \text{if}\, \mathsecret{\prot{fake}}\, \text{else}\, \mathsecret{\prot{fZero}}$ \text{obliviously}\;
        \comm{\cfont{Update \secret{\prot{Used}}}}
        $\code{ORAM}.\mathsecret{\prot{Used}} = \prot{OR}(\mathsecret{\prot{mask}}, \code{ORAM}.\mathsecret{\prot{Used}})$\;
        \comm{\cfont{Get the corresponding index}}
        $\mathsecret{\prot{index}} \leftarrow \prot{DOT}(\code{ORAM}.\mathsecret{\prot{Data}}, \mathsecret{\prot{mask}})$ \;
        Reveal $ p \leftarrow \prot{index}$ in plaintext \;
        \Return{$p$;}
\end{algorithm}

\begin{algorithm}[t]
	\small
        \SetKwInOut{Global}{\textbf{Global}}
	\SetKwInOut{Input}{\textbf{Inputs}}
	\SetKwInOut{Output}{\textbf{Output}}
	\caption{\small \code{UniqueNeighborsCount}}\label{algo:unique_neighbors_count}
	\Input{Target vertex $\mathsecret{v}$ and the target block ID $\mathsecret{i} = \mathsecret{\lceil \frac{v}{k} \rceil}$.}
	\Output{$\mathsecret{{\sf{num}}}^{A}$, the number of $v$'s unique outing neighbors. }
	% \BlankLine
        \comm{\cfont{Sub-graph extraction. }}
        % \BlankLine
        Fetch the target edge blocks $\mathsecret{B} \leftarrow \code{VORAM}.{\sf{access}}(\mathsecret{i})$, where $\mathsecret{B}$ contains $(bl)$ $\sf{source\_nodes}$ and $\sf{dest\_nodes}$\;
        % \BlankLine
        \comm{\cfont{Parallely sub-graph process. }}
        % \BlankLine
        \comm{\cfont{1) Mask-out the non-neighbors.}}
        Construct $\mathsecret{\vec{v}}$ by expanding $\mathsecret{v}$ $bl$ times\;
        Compute $\mathsecret{{\sf{mask}}} \leftarrow {\sf{EQ}}(\mathsecret{\vec{v_s}}, \mathsecret{B}.{\sf{source\_nodes}})$ \;
        Compute $\mathsecret{{\sf{candidate}}} \leftarrow {\sf{MUL}}(\mathsecret{{\sf{mask}}}, \mathsecret{B}.{\sf{dest\_nodes}})$ \;
        \comm{\cfont{2) De-duplicate neighbors.}}
        $\mathsecret{{\sf{same\_mask}}} \leftarrow {\sf{EQ}}(\mathsecret{{\sf{candidate}}}_{[1:]}, \mathsecret{{\sf{candidate}}}_{[:-1]})$\;
        $\mathsecret{{\sf{same\_mask}}}.{\sf{append}(\mathsecret{1})}$\;
        $\mathsecret{\sf{mask}} \leftarrow {\sf{MUL}}(\mathsecret{{\sf{same\_mask}}}, \mathsecret{{\sf{mask}}})$\;
        \comm{\cfont{3) Aggregating for the final outcomes.}}
        Compute $\mathsecret{\sf{mask}}^{A} \leftarrow {\sf{B2A}(\mathsecret{{\sf{mask}}})}$ \;
        % \BlankLine
        \While{$\lceil \frac{l}{2} \rceil \ge 1$}{
            Pads $\mathsecret{0}^{A}$ to $\mathsecret{\sf{mask}}^{A}$ to be even \;
            Split $\mathsecret{\sf{mask}}^{A}$ half-by-half to $\mathsecret{\sf{mask}}^{A}_l$ and $\mathsecret{\sf{mask}}^{A}_r$\;
            Aggregate $\mathsecret{\sf{mask}}^{A} \leftarrow {\sf{ADD}}(\mathsecret{\sf{mask}}^{A}_l, \mathsecret{\sf{mask}}^{A}_r)$ \;
            $l = {\sf{len}}(\mathsecret{{\sf{mask}}}^{A}) / 2$ \;
        }
        $\mathsecret{{\sf{num}}}^{A} = \mathsecret{\sf{mask}}^{A}$\;
        \Return{$\mathsecret{{\sf{num}}}^{A}$;}
\end{algorithm}

\section{Prefix-based ORAM.{\sf{Access}}}\label{app:oram_access}

The original Square-root ORAM~\cite{zahur2016revisiting} uses a $O(n)$ rounds method in the last level of its recursive ORAM (\code{GetPosBase} function, in Section D, Figure 6), and in \sysname, we optimize it to $O(\log(n))$ rounds. The interfaces and the methods are shown in Algorithm~\ref{algo:prefix-access}, keeping the same notations as the original Square-root ORAM. 

The complicated part in Square-root ORAM is the extraction of the first unused element in the last level ORAM obliviously as the used elements are defined by users access patterns. Algorithm~\ref{algo:prefix-access} locates this information by constructing $\mathsecret{\prot{fZero}}$ (first unused element) leveraging the prefix-computations (Lines 1-6), propagating \emph{whether there exist 1 in $\mathsecret{\prot{notUsed}}$ before my location, including my location} to the following elements obliviously. 
After line 6, $\mathsecret{\prot{fZero}}$ contains a successive $\mathsecret{0}$ and follows with $\mathsecret{1}$, and the last $\mathsecret{1}$ indicates the element which is the first unused element, \ie the first zero in $\mathsecret{\prot{Used}}$, the first one in $\mathsecret{\prot{notUsed}}$.
Then, we obliviously transfers the previous $\mathsecret{1}$ to $\mathsecret{0}$ by a differential $\prot{XOR}$ (Lines 7-8). Note that $\mathsecret{\prot{fZero}}$ corresponds to the \secret{s_2} while not considering \secret{\prot{fake}} in the original Square-root ORAM. 

Lines 9 corresponds to $\mathsecret{s_1}$ of Square-root ORAM, indicating which element corresponding to the cipher index $\mathsecret{i}$. Note that before Line 10, there are only $\mathsecret{1}$ in $\mathsecret{\prot{fZero}}$ and $\mathsecret{s_1}$ and the $\mathsecret{1}$ indicates the expected element for $\mathsecret{\prot{fake}}$ is false or true use cases. We obliviously select $\mathsecret{\prot{fZero}}$ or $\mathsecret{s_1}$ based on $\mathsecret{\prot{fake}}$ and obtain the $\mathsecret{\prot{mask}}$. Note that the $\mathsecret{1}$ in $\mathsecret{\prot{mask}}$ indicates the to-be-use elements location, therefore we update \code{ORAM}.$\mathsecret{\prot{Used}}$ afterwards (Line 11). Then, we get the corresponding element in \code{ORAM}.$\mathsecret{\prot{Data}}$ using dot-product, \ie only the elements corresponding to $\mathsecret{1}$ of $\mathsecret{\prot{mask}}$ is preserved, which is the expected physical index.

\section{Other Queries}\label{app:other_queries}
We present the implementation of \code{UniqueNeighborsCount} in Algorithm~\ref{algo:unique_neighbors_count}, which adds a de-duplication phase between the sub-graph extraction and aggregation phases, updating the $\mathsecret{\prot{mask}}$ eliminating the duplicate neighbors (Lines 2-7). Specifically, the de-duplication procedure is similar to \code{NeighborsGet} without extracting the real neighbors.

\balance

\end{document}